\documentclass{PoS}
\usepackage{amsmath}
\usepackage{amssymb}
\title{Energy Dependence of Multiplicity Fluctuations\\
 in Heavy Ion Collisions}

\ShortTitle{Energy Dependence of Multiplicity Fluctuations}

\author{\speaker{Benjamin Lungwitz}, 
        Fachbereich Physik der Universit\"{a}t, Frankfurt, Germany.\\
        E-mail: \email{lungwitz@ikf.uni-frankfurt.de}}

\author{for the NA49 Collaboration}

\author{
\vspace{0.5cm}
C.~Alt$^{9}$, T.~Anticic$^{23}$, B.~Baatar$^{8}$,D.~Barna$^{4}$,
J.~Bartke$^{6}$, L.~Betev$^{10}$, H.~Bia{\l}\-kowska$^{20}$,
C.~Blume$^{9}$,  B.~Boimska$^{20}$, M.~Botje$^{1}$
% ...
\\

\vspace{0.5cm}
\noindent
$^{1}$NIKHEF, Amsterdam, Netherlands. \\
% ...
$^{4}$KFKI Research Institute for Particle and Nuclear Physics, Budapest, Hungary.\\
% ...
$^{6}$Institute of Nuclear Physics, Cracow, Poland.\\
% ...
$^{8}$Joint Institute for Nuclear Research, Dubna, Russia.\\
$^{9}$Fachbereich Physik der Universit\"{a}t, Frankfurt, Germany.\\
$^{10}$CERN, Geneva, Switzerland.\\
% ...
$^{20}$Institute for Nuclear Studies, Warsaw, Poland.\\
% ...
$^{23}$Rudjer Boskovic Institute, Zagreb, Croatia.\\

}

\abstract{
The energy dependence of multiplicity fluctuations was studied for the most central $Pb+Pb$ collisions at $20A$, $30A$, $40A$, $80A$ and 
$158A$ GeV by the NA49 experiment at the CERN SPS.
The multiplicity distribution for negatively and positively charged hadrons is significantly narrower than
Poisson one for all energies.
No significant structure in energy dependence of the scaled variance of multiplicity fluctuations is observed. The measured scaled variance
is lower than the one predicted by the grand-canonical formulation of the hadron-resonance gas model. The results for scaled variance
are in approximate agreement 
with the string-hadronic model UrQMD.
}

\FullConference{Correlations and Fluctuations in Relativistic Nuclear Collisions\\
		 July 7-9 2006\\
		 Florence, Italy}

\begin{document}

\section{Introduction}
At high energy densities ($\approx 1\, GeV/fm^3$) a phase transition from hadron gas to quark-gluon-plasma (QGP) is expected to occur.
There are indications that at RHIC and top SPS energies quark-gluon-plasma is created at the early stage of heavy ion collisions~\cite{Heinz:2000bk, Gyulassy:2004zy}.
The energy dependence of 
various observables show anomalies at low SPS energies which might be related to the onset of deconfinement~\cite{Gazdzicki:2004ef}. 
Lattice QCD calculations suggest furthermore the existence of a critical point in the phase diagram of strongly interacting matter which separates the 
line of the first order phase transition from a crossover. Models predict an increase of multiplicity fluctuations near the onset of deconfinement~\cite{Gazdzicki:2003bb} 
or the critical point~\cite{Stephanov:1999zu}.

This motivates a vigorous experimental and theoretical study of multiplicity fluctuations in high energy nuclear 
collisions~\cite{Rybczynski:2004yw, Aggarwal:2001aa}.
In addition to the effects mentioned above also the ``background'' statistical fluctuations are interesting to study.
In a grand-canonical 
statistical model scaled variance $\omega$\footnote{
a commonly used measure of fluctuations, $\omega=\frac{Var(n)}{<n>}$, where $Var(n)$ and $<n>$ are variance and mean of multiplicity distributions, respectively.}
is close to one. In the past it was commonly believed that the result of the grand-canonical, canonical and micro-canonical ensemble should be similar in
the infinite volume limit, like it is for mean quantities. Recent calculations showed that this assumption is wrong. The scaled variance in the
infinite volume limit is largest for the grand-canonical ensemble and smallest for the micro-canonical one~\cite{Begun:2004gs, Begun:2004pk}. It is because the fluctuations in the
canonical and micro-canonical ensembles are reduced by conservation laws and this reduction is, unlike for the mean, volume independent for $V\rightarrow \infty$.

The first preliminary results on the energy dependence of multiplicity fluctuations in central $Pb+Pb$ collisions at SPS energies 
were presented in this paper in order
to look for experimental signatures of increased fluctuations due to a phase transition or the critical point and for the predicted reduction of fluctuations
in relativistic hadron gas due to conservation laws.

\section{The NA49 Experiment}

\begin{figure}
\includegraphics[height=14cm,angle=270]{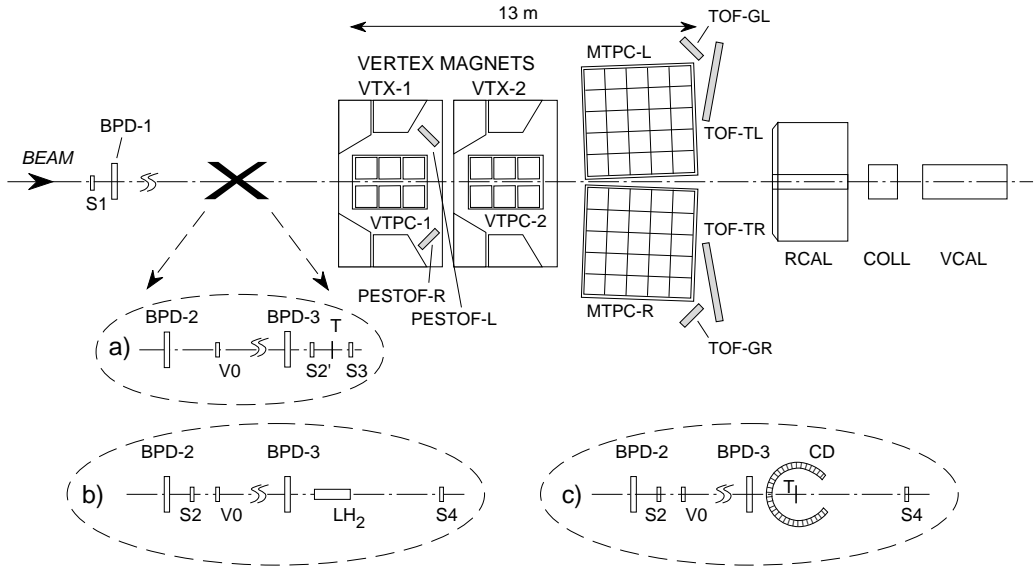}
\caption{\label{na49_setup}Setup of the NA49 experiment and different target configurations, a) for $Pb+Pb$, b) for $p+p$
and c) for $p+A$ collisions.}
\end{figure}
The NA49 detector~\cite{Afanasev:1999iu} (see figure~\ref{na49_setup})
is a large acceptance fixed target hadron spectrometer. Its main devices are four large 
time projection chambers (TPCs).
Two of them, called vertex TPCs, are located in two superconducting dipole magnets with a total bending power of $7.8$ Tm. 
The other two TPCs are installed behind the magnets left and
right of the beam line allowing precise particle tracking in the high density region of heavy ion collisions.
The measurement of the energy loss $dE/dx$ allows particle identification and a good rejection of electrons in a large momentum regime.

A large variety of different colliding systems was studied. Beams of p and Pb are available at the CERN SPS directly, C and Si beams 
were produced via fragmentation of the primary Pb beam. Targets of liquid hydrogen or solid foils of different materials were used.
Data on central $Pb+Pb$ collisions at $20A$, $30A$, $40A$, $80A$ and $158A$ GeV were recorded and analysed for this analysis.

\subsection{Centrality Determination}\label{centr_det}

The downstream veto calorimeter, originally designed for NA5,  allows a determination of the centrality of a collision by measuring the energy in the 
projectile spectator region~\cite{Appelshauser:1998tt}.
%This measure of centrality is mostly independent of produced particles which is important in a study of multiplicity fluctuations.
\begin{figure}
\centerline{\includegraphics[height=6cm]{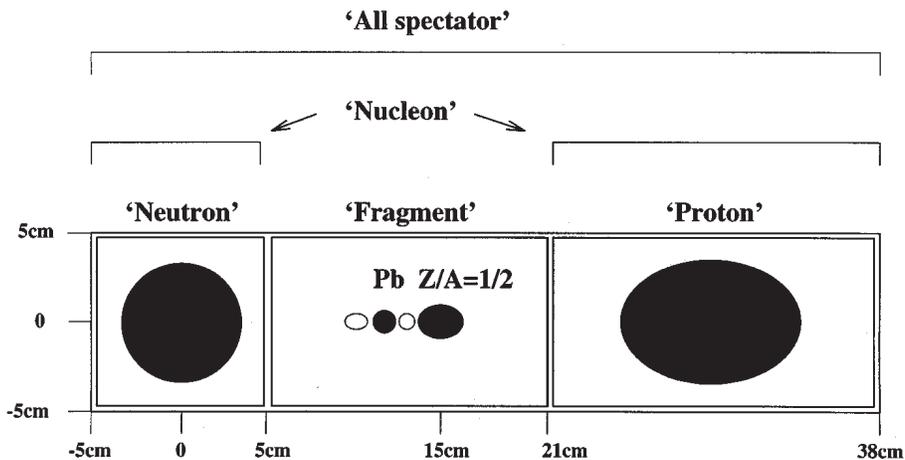}}
\caption{\label{collimator}Collimator in front of the Veto calorimeter at $158A$ GeV~\cite{Appelshauser:1998tt}.}
\end{figure}
A collimator is located $25$~m downstream from the target and is adjusted for each energy in such a way that all projectile spectator protons, 
neutrons and
fragments can reach the veto calorimeter. For $158A$ GeV the hole in the collimator is with respect to the beam axis $\pm 5$~cm in vertical direction
and $-5$~cm and $+38$~cm in horizontal direction taking into account the deflection of charged particles by the magnetic field (figure~\ref{collimator}).
For $40A$ and $80A$ GeV the hole is $\pm 12$~cm in vertical and $-13$~cm respectively $+47$~cm in horizontal direction.
For $20A$ and $30A$ GeV the ring calorimeter with a hole of radius $28$~cm and positioned $18$~m away from the target and $10$~cm away from beam axis in
horizontal direction is used as a collimator.
For higher energies the ring calorimeter is positioned $17$~cm off in horizontal direction and has only a small influence on the acceptance of 
the veto calorimeter.

\begin{figure}
\includegraphics[width=7cm]{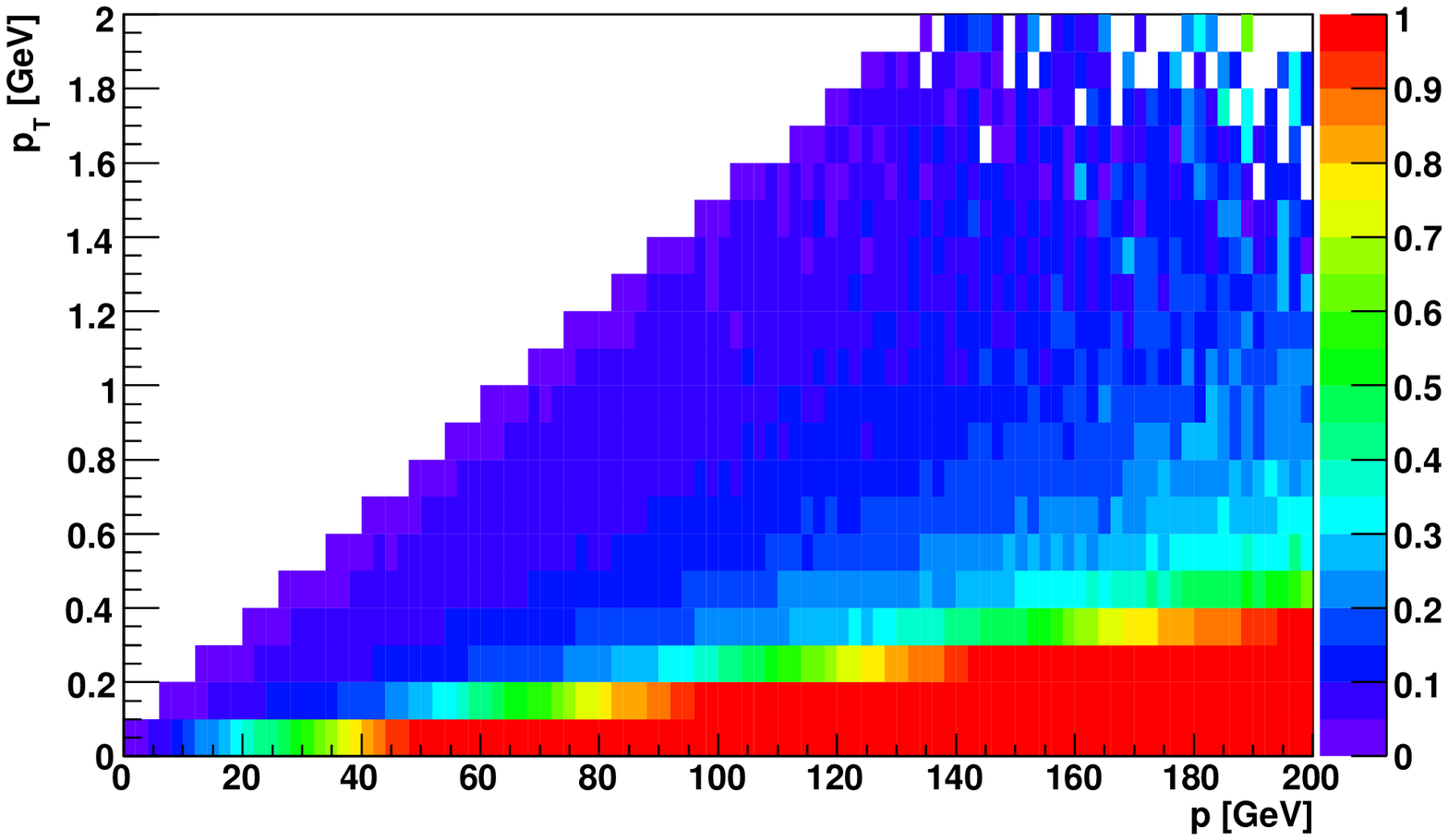}
\includegraphics[width=7cm]{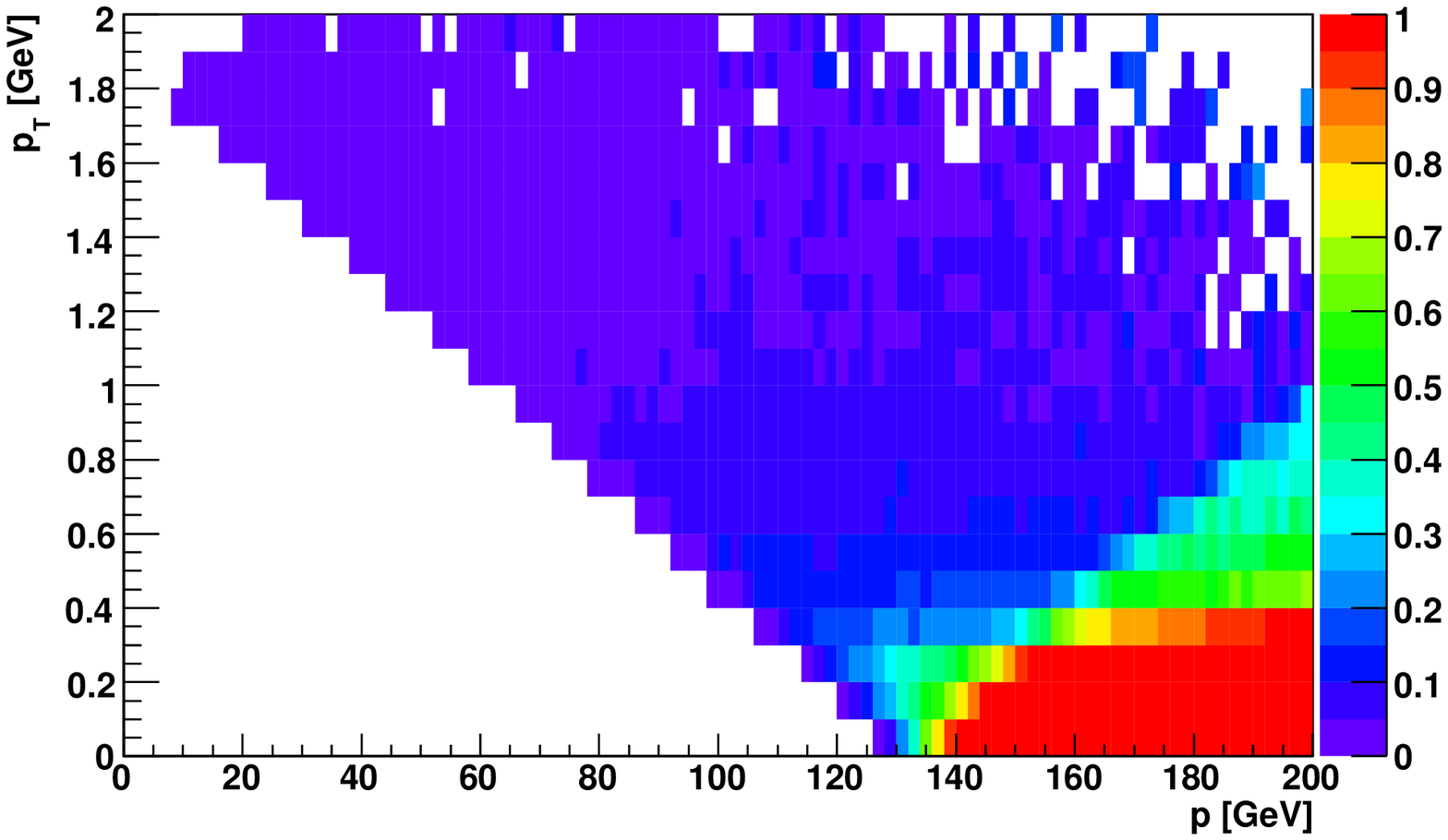}
\caption{\label{veto_acc}Acceptance of the Veto calorimeter for neutral (left) and positively charged (right) main vertex particles at $158A$ GeV as a function of total 
momentum $p$ and transverse momentum $p_T$.}
\end{figure}

For the following analysis the collisions are selected according to their energy in the veto calorimeter. 
A large fraction of the Veto energy is due to projectile spectators.
However also a fraction of non-spectator particles, mostly 
protons and neutrons, contribute to Veto energy. The acceptance of the Veto calorimeter at $158A$ GeV is shown in figure~\ref{veto_acc} for neutral and
positively charged particles.

The centrality $C$ of an event with a veto energy $E_{Veto}$ can be calculated using the known trigger centrality 
$C_{trig}=\frac{\sigma_{trig}}{\sigma_{inel}}$ and the veto energy distribution as:
\begin{equation}
C=\frac{\sigma_{E_{Veto}}}{\sigma_{inel}}=C_{trig} \cdot \frac{\int_{0}^{E_{Veto}}{dN/dE_{Veto,trig}} }{\int_{0}^{\infty} {dN/dE_{Veto,trig}}}
\end{equation}
where $dN/dE_{Veto,trig}$ is the Veto energy distribution for a given trigger.

\subsection{Track Selection}

\begin{figure}
\includegraphics[width=8.3cm]{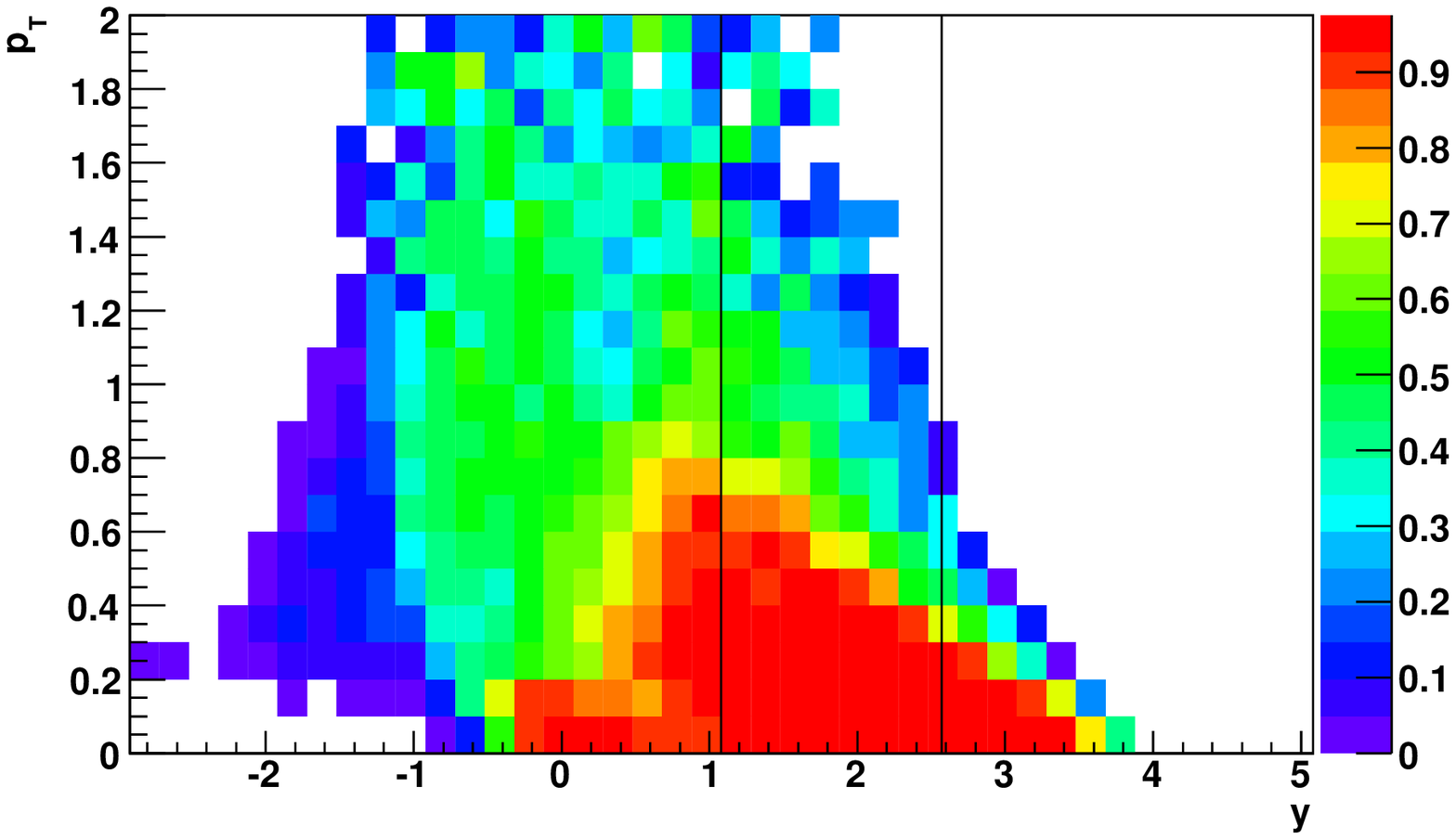}
\includegraphics[width=7cm]{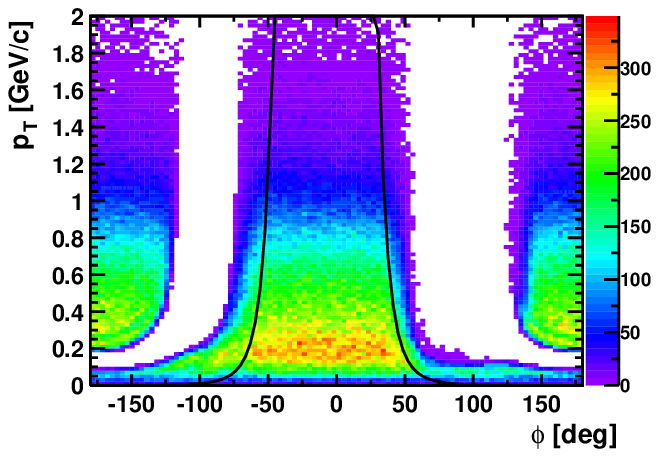}
\caption{\label{acceptance_Pb160}Acceptance for negative hadrons at $158A$ GeV as a function of rapidity (pion mass assumed) and $p_T$ (left), 
as well as a function of $\phi$ and $p_T$ ($1.4 < y < 1.6$, right).}
\end{figure}

\begin{figure}
\centerline{\includegraphics[width=10cm]{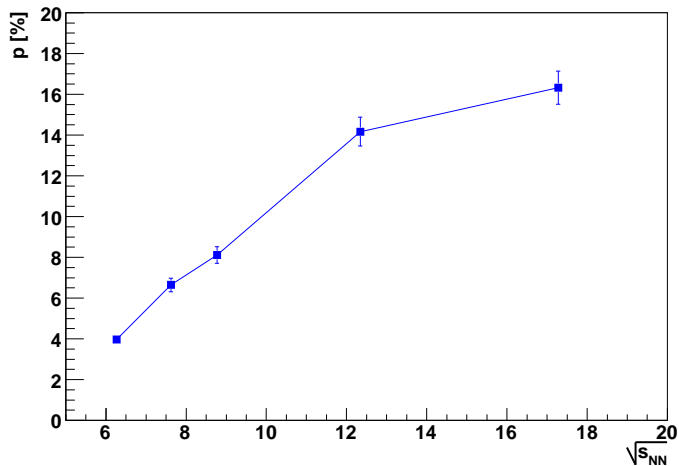}}
\caption{\label{acceptance}Acceptance for negative hadrons as a function of collision energy calculated using Venus model and GEANT detector simulation.
The acceptance was also estimated using the ratio of measured multiplicity of negatively charged hadrons to the published or preliminary NA49 total
$\pi^-$ and $K^-$ yields, the difference ($<5\%$) is shown by vertical error bars.}
\end{figure}

Since detector effects like track reconstruction efficiency might have a large influence on multiplicity fluctuations, it is important to select a 
very clean track sample for the
analysis (figure~\ref{acceptance_Pb160}). Therefore the acceptance for this analysis is limited to a part of the forward hemisphere, where the NA49 detector has the
highest tracking efficiency.
This was done by restricting the analysis to the rapidity interval $1<y(\pi)<y_{beam}$
\footnote{Rapidity is calculated in the center of mass system assuming pion mass.}
 for $20A$ to $80A$ GeV and $1.08<y(\pi)<2.57$ for $158A$ GeV\footnote{A different rapidity cut
is used for this energy because of missing geometrical acceptance near beam rapidity~\cite{Rybczynski}.}. In addition a cut on transverse momentum according 
to \cite{Alt:2004ir} was applied, which is dependent both on rapidity and azimuthal angle.
This ensures that only tracks with sufficient points both in the vertex TPC 2 and in one of the main TPCs are used.
For them the reconstruction efficiency is larger than $98\%$. The disadvantage of this track selection is that the fraction of accepted tracks 
is small and depends on collision energy, see figure~\ref{acceptance}.

A cut on the distance between the interaction point and the track trajectory extrapolated to the target is used to reduce the background of secondary particles 
originating from weak decays (track impact parameter)\footnote{The difference has to be smaller than $4$~cm orthogonal and $2$~cm parallel to the magnetic field direction.}. 
Removing this cut would change the scaled  variance by less than $2 \%$.\\
A cut on the energy loss of particles in the detector gas ($dE/dx$) removes electrons and allows the study of hadron multiplicity fluctuations only. 
The scaled variance will be less than $2\%$ different if this cut is removed. 

\section{Fluctuation Analysis}

\begin{figure}
\begin{center}
\includegraphics[height=4cm]{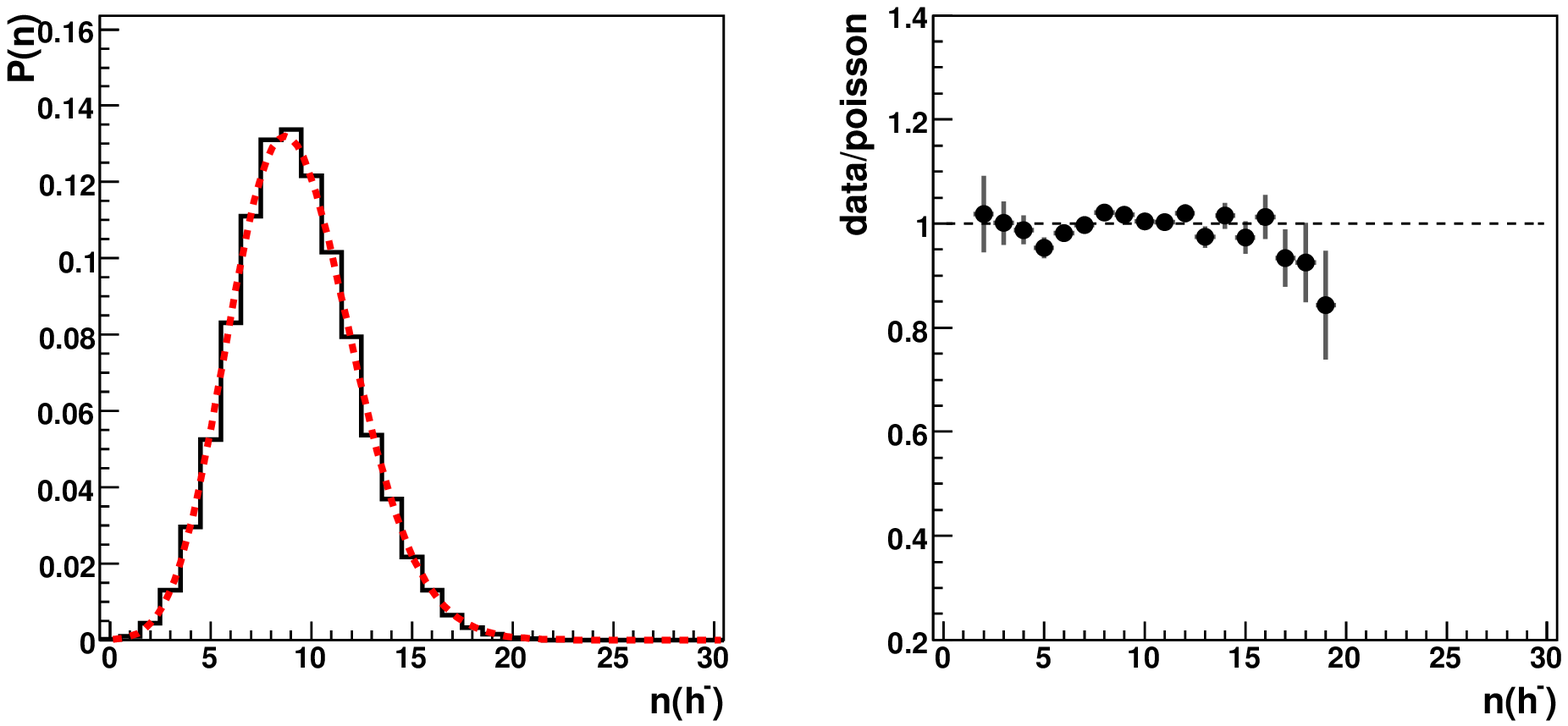}\\
\includegraphics[height=4cm]{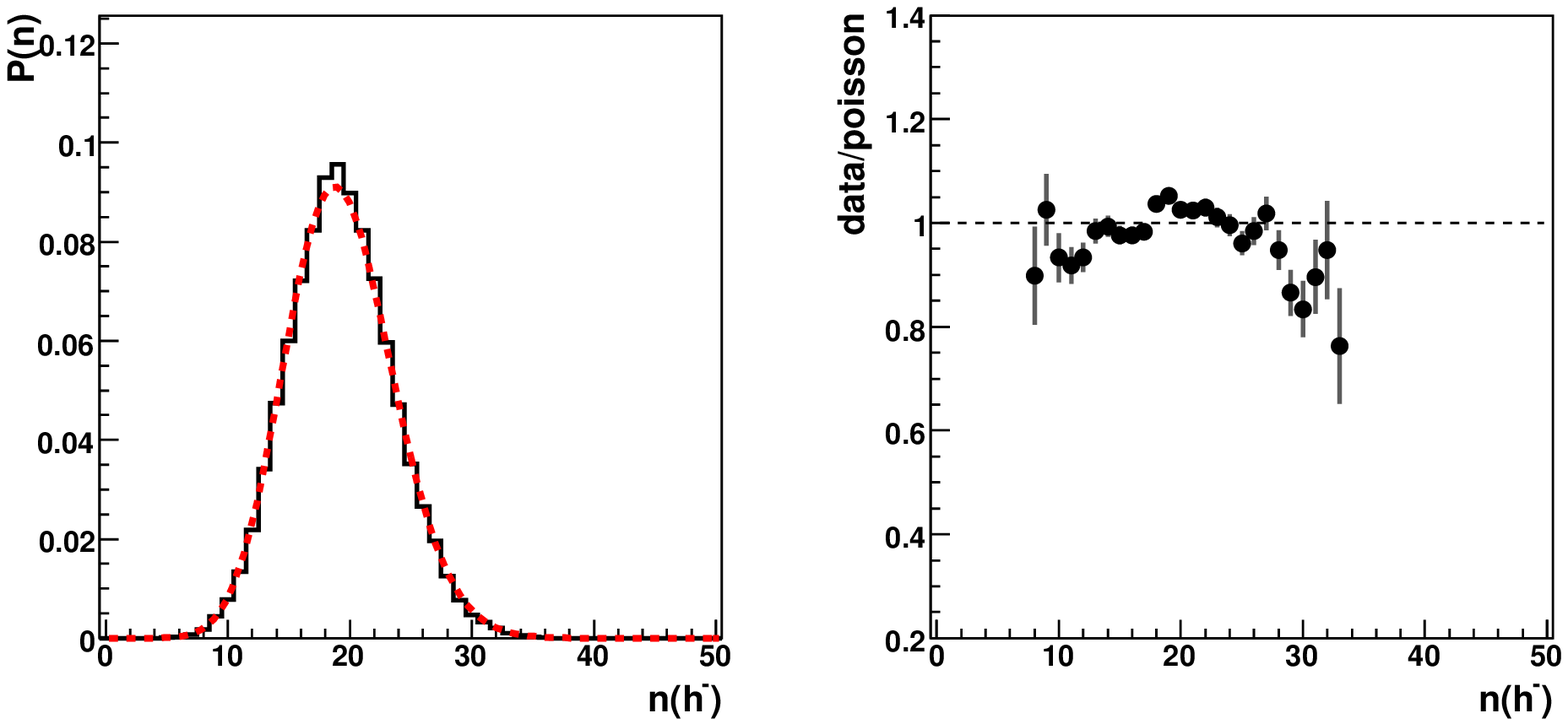}\\
\includegraphics[height=4cm]{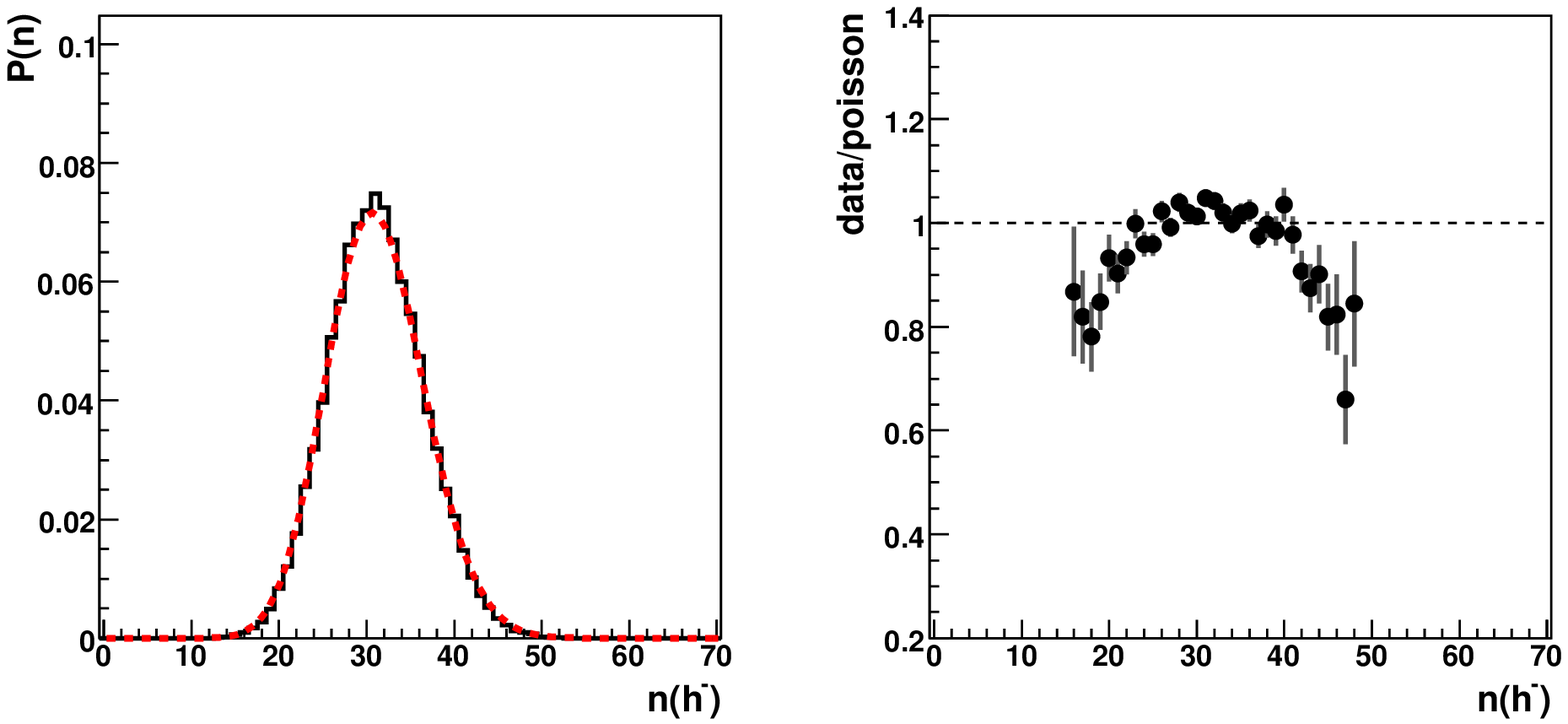}\\
\includegraphics[height=4cm]{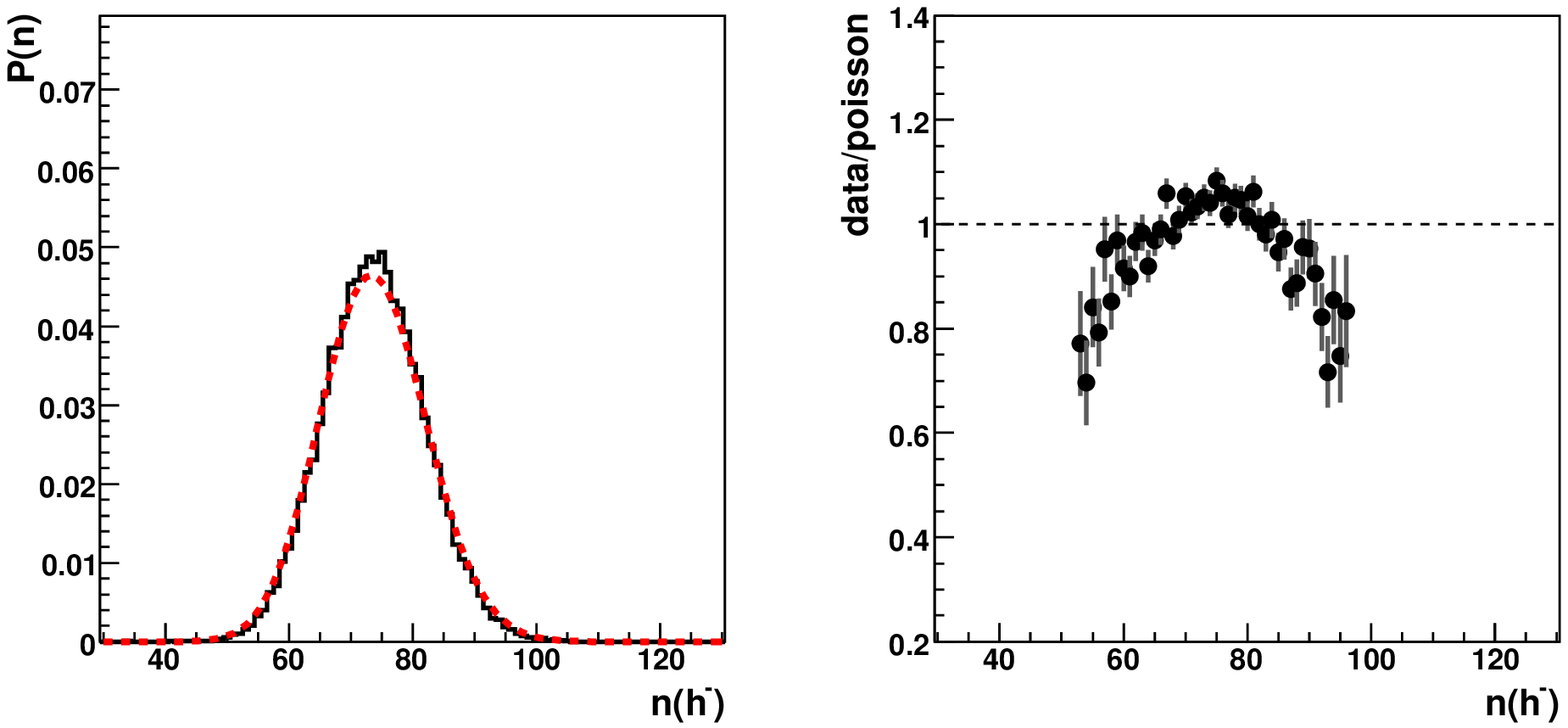}\\
\includegraphics[height=4cm]{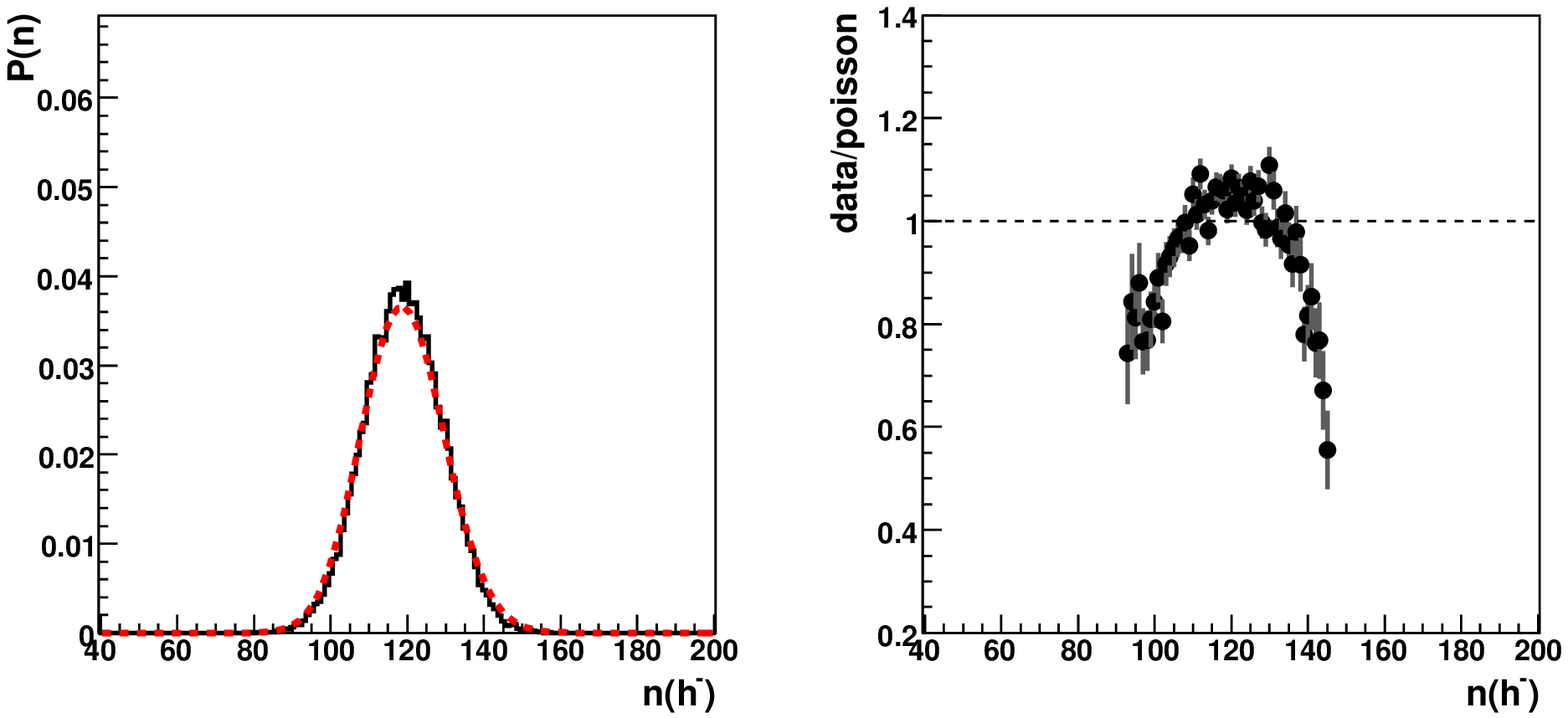}
\end{center}
\caption{\label{mult_dist}Left: multiplicity distribution of negative hadrons in central ($C<1\%$) $Pb+Pb$ collisions from $20A$ (top) to  $158A$ GeV (bottom).
The red line indicates a Poisson distribution with the same mean multiplicity. Right: ratio of the measured multiplicity distribution over Poisson distribution.
It is visible that the measured distribution is significantly narrower than the Poissonian one for all energies.
Note that the multiplicity fluctuations are not corrected for the finite centrality bin width.}
\end{figure}

The multiplicity distribution in central ($7-10\%$) $Pb+Pb$ collisions at $158A$ GeV is shown in figure~\ref{mult_dist}.
The measured distribution is significantly narrower than a Poisson one.\\ 
The basic measure of multiplicity fluctuations used in this analysis is the scaled variance:
\begin{equation}
\omega=\frac{Var(n)}{<n>}
\end{equation}
where $Var(n)$ and $<n>$ are variance and mean of multiplicity distributions, respectively.

The scaled variance for positively ($\omega(h^+)$), negatively ($\omega(h^-)$) and all charged hadrons ($\omega(h^\pm)$) will be presented.
The scaled variance of a Poisson distribution is 1, independent of its mean multiplicity. A larger $\omega$ might indicate additional non-statistical fluctuations,
a smaller $\omega$ might be a hint for a suppression of fluctuations e.g. due to conservation laws.

\begin{figure}
\includegraphics[height=7.1cm]{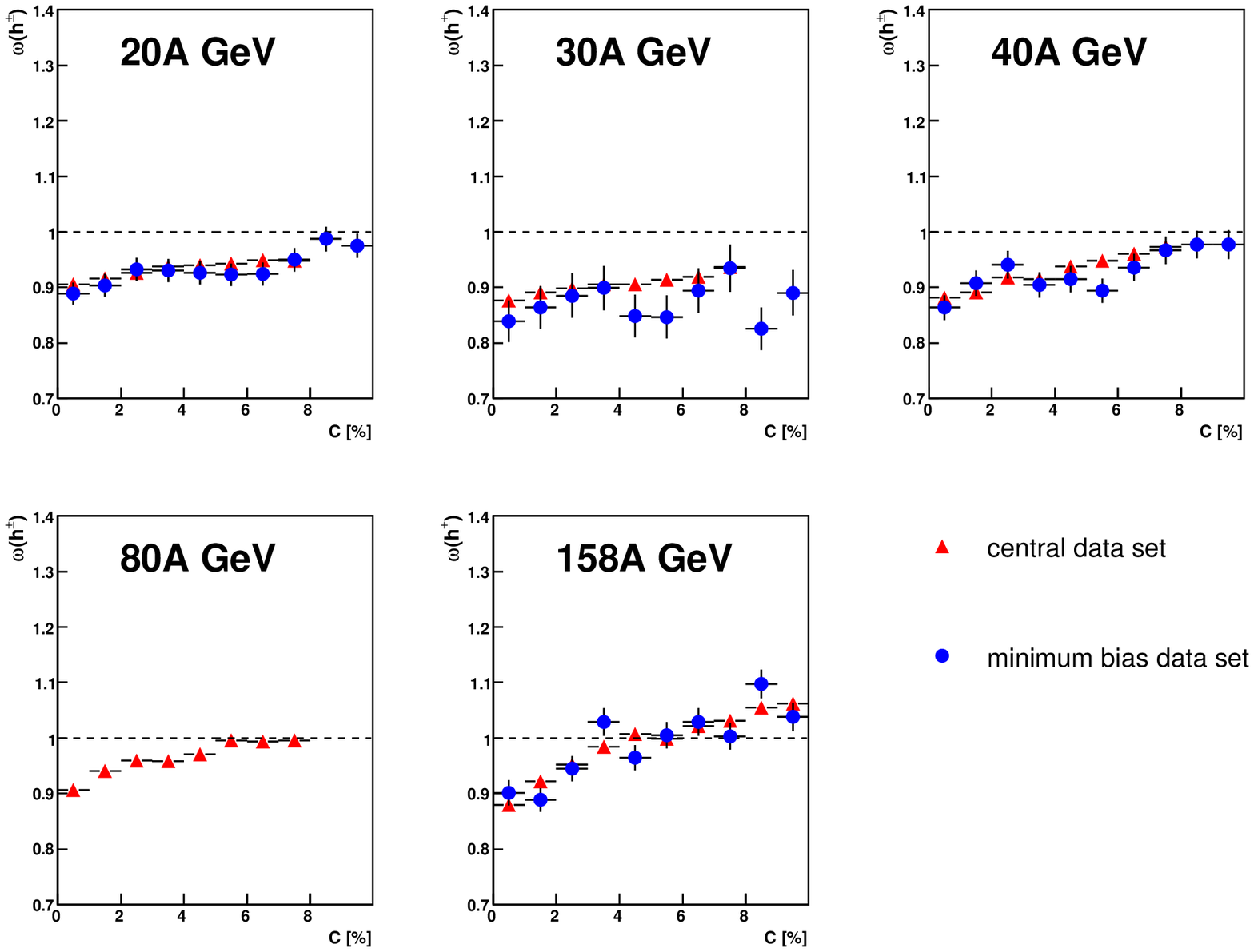}\\
\includegraphics[height=7.1cm]{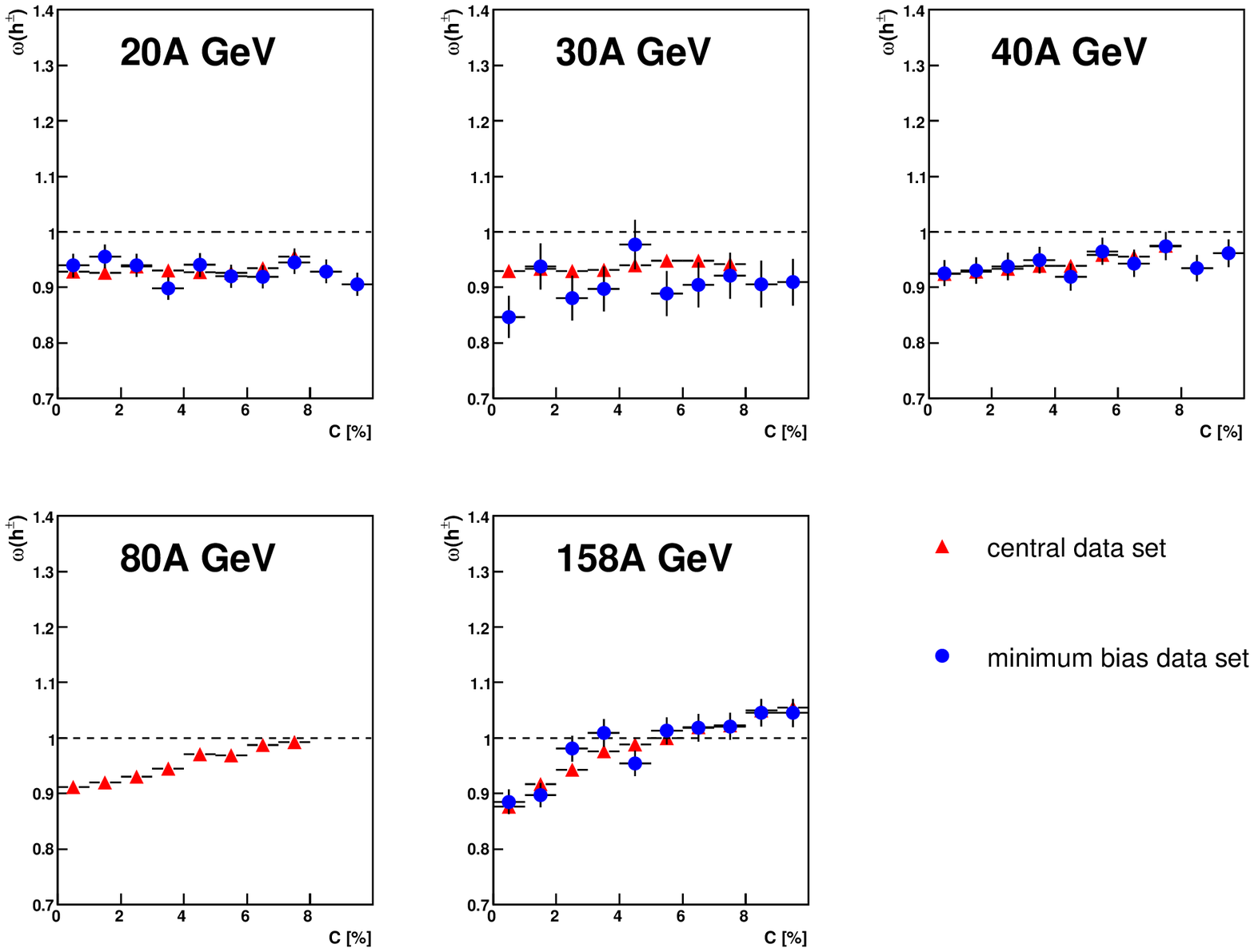}\\
\includegraphics[height=7.1cm]{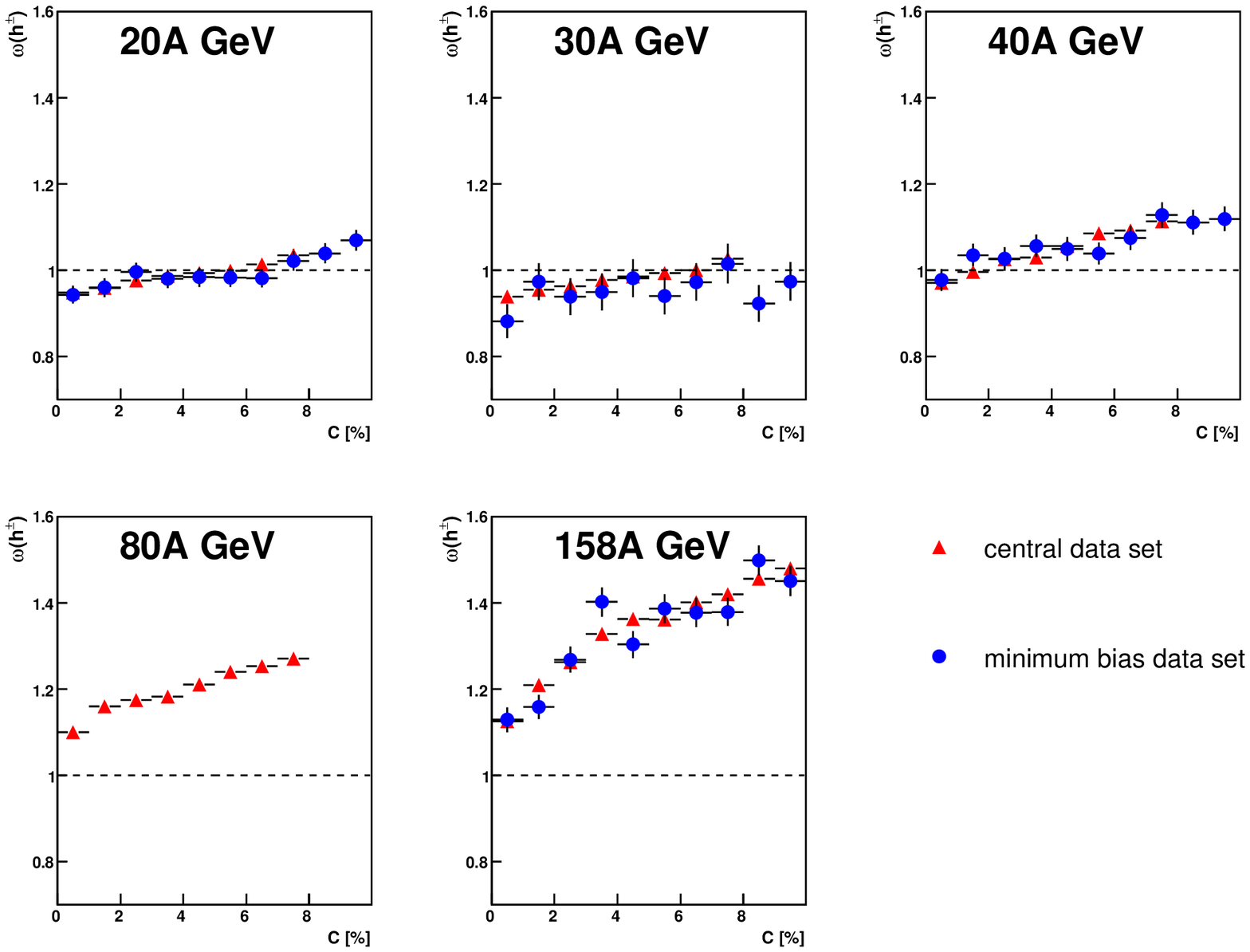}
\caption{\label{centrdep_hm}Centrality dependence of $\omega(h^+)$ (top), $\omega(h^-)$ (middle) and  $\omega(h^\pm)$ (bottom) for $Pb+Pb$ collisions 
at different energies. $C<1\%$ corresponds to the most central collisions. The shown centrality range of $C<10\%$ corresponds
approximately to number of projectile participants $N_P^{proj}>160$. Only statistical errors are shown.}
\end{figure}

Figure~\ref{centrdep_hm} shows the centrality dependence of multiplicity fluctuations in $Pb+Pb$ collisions at different energies for central collisions.
In general $\omega$ decreases with increasing centrality (i.e. decreasing Veto energy), this trend is stronger for higher energies.

The results for collisions selected by minimum bias and central on-line trigger are shown separately in figure~\ref{centrdep_hm} and they agree within errors.
For further analysis the results for central data sets are
used because of larger statistics in comparison to the minimum bias data sets.

\subsection{Experimental Biases}

In superposition models (e.g. the Wounded Nucleon Model~\cite{Bialas:1976ed}) the total number of particles produced in a collision $n$ is the sum of the 
number of particles 
produced by different independent particle production sources:
\begin{equation}
n=\sum_i{n_{i,so}}
\end{equation}
In these models the scaled variance has two contributions. The first is due to the fluctuations of the number of particles emitted by a single source $\omega_{so}$, 
the second
is due to the fluctuations in the number of sources $\omega_k$:
\begin{equation}
\omega=\omega_{so}+<n_{so}>\cdot \omega_k
\end{equation}
where $<n_{so}>$ is the mean multiplicity of hadrons from a single source.\\
The number of sources related to projectile participants is fixed by fixing the
energy detected in the Veto Calorimeter and consequently the number of projectile spectators (see chapter~\ref{centr_det}). 

For very central collisions analysed here the Veto energy is dominated by the energy of forward going participants and not spectators. 
For these collisions
the number of projectile participants is in good approximation fixed and independent of Veto energy. Therefore no correction for the finite width
of the Veto energy bins is applied.

The finite resolution of the Veto calorimeter introduces additional fluctuations in $N_P^{proj}$ even for a fixed Veto energy.
For very central collisions their influence on the scaled variance is estimated to be small, in contrast to peripheral collisions. 
The data is not corrected for this effect, introducing a correction would
reduce the scaled variance by less than $5\%$.

\begin{figure}
\centerline{\includegraphics[height=8cm]{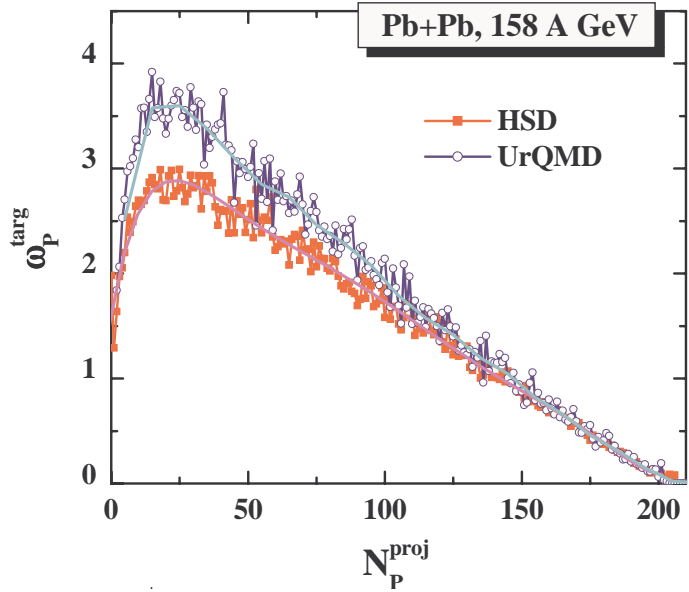}}
\caption{\label{nptarg_fluct}Fluctuations in the number of target participants for a fixed number of projectile participants in the UrQMD and HSD~\cite{Konchakovski:2005hq}
models.}
\end{figure}

In the NA49 experiment the number of target participants $N_P^{targ}$ is not fixed, it fluctuates substantially (see figure~\ref{nptarg_fluct}). 
These fluctuations might contribute even to multiplicity fluctuations in the forward hemisphere and increase the scaled variance~\cite{Rybczynski,Gazdzicki:2005rr}.\\
Model calculations (HSD, UrQMD)~\cite{Konchakovski:2005hq} show that the scaled variance of target participant fluctuations 
$\omega(N_P^{targ})=\frac{Var(N_P^{targ})}{<N_P^{targ}>}$ is
small for central collisions, and thus the corresponding increase of the scaled variance of multiplicity distributions is also expected to be small.

In order to minimize the contribution due to both Veto calorimeter resolution and $N_P^{targ}$ fluctuations further analysis was done for very central
collisions ($C<1\%$, which corresponds to $N_P^{proj}>193$).

The systematic error is estimated taking into account the influence of the track impact parameter cut ($\pm 2\%$), the cut on electron rejection ($\pm 2\%$),
the effect of finite resolution of the Veto calorimeter ($-5\%$) and 
possible remaining participant fluctuations due to the finite width of the selected centrality bin ($-5\%$). 
The total systematic error is estimated to be $\sigma_{sys}\approx^{+2}_{-5}\%$.

Model calculations (Venus and UrQMD) show that for higher energies ($80A$ and $158A$ GeV) the details of centrality selection have a significant effect on
multiplicity fluctuations. When the centrality is determined by fixing the number of projectile spectators the scaled variance obtained by these models is up to
$30\%$ larger than the one obtained by fixing the energy in the Veto Calorimeter using a detailed simulation of its acceptance.
For lower energies ($20A$- $40A$ GeV) no significant difference of the scaled variance for different centrality selections was observed.

\section{Results and Discussion}\label{model_comp}

\begin{figure}
\includegraphics[height=8cm]{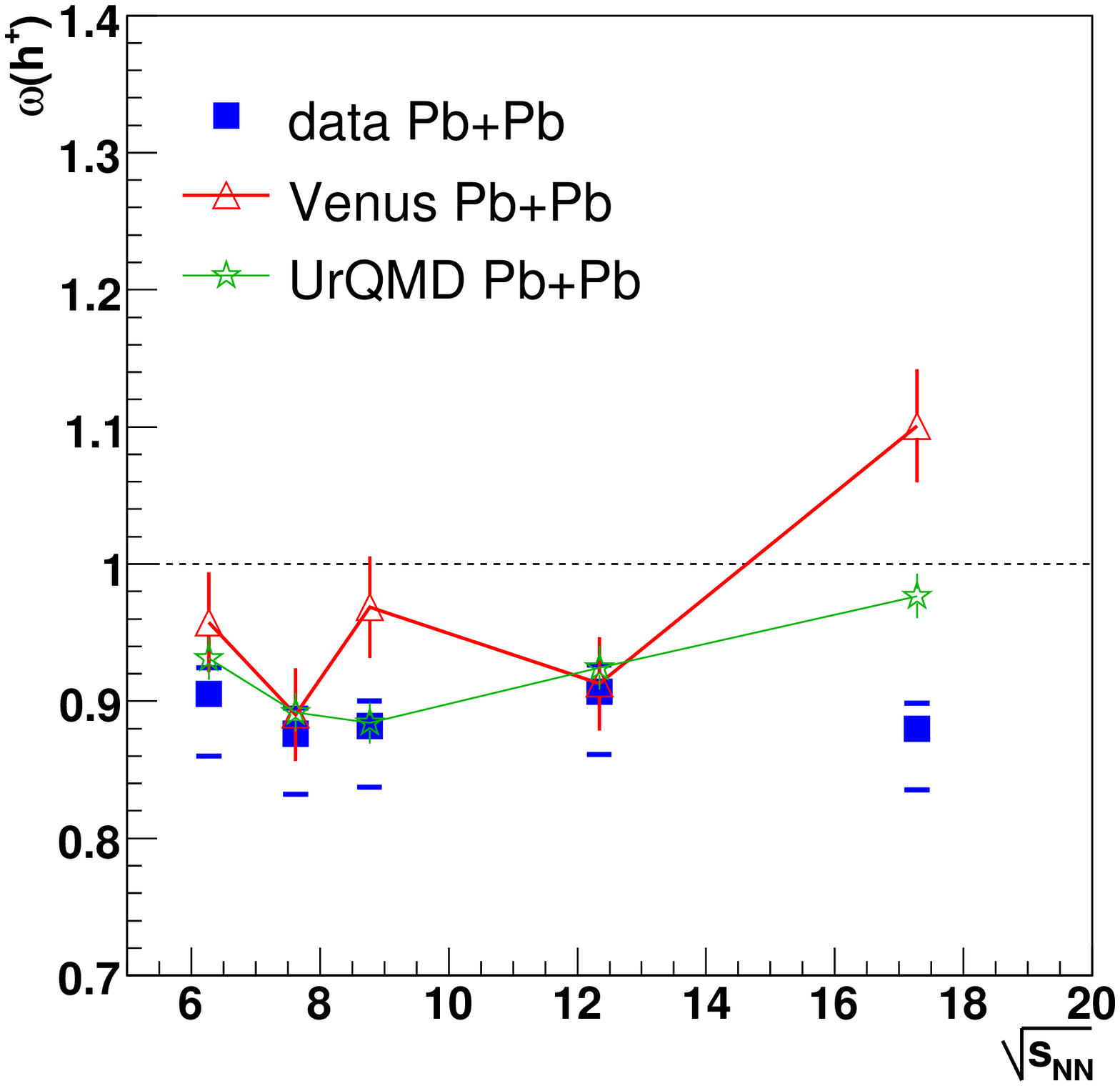}
\includegraphics[height=8cm]{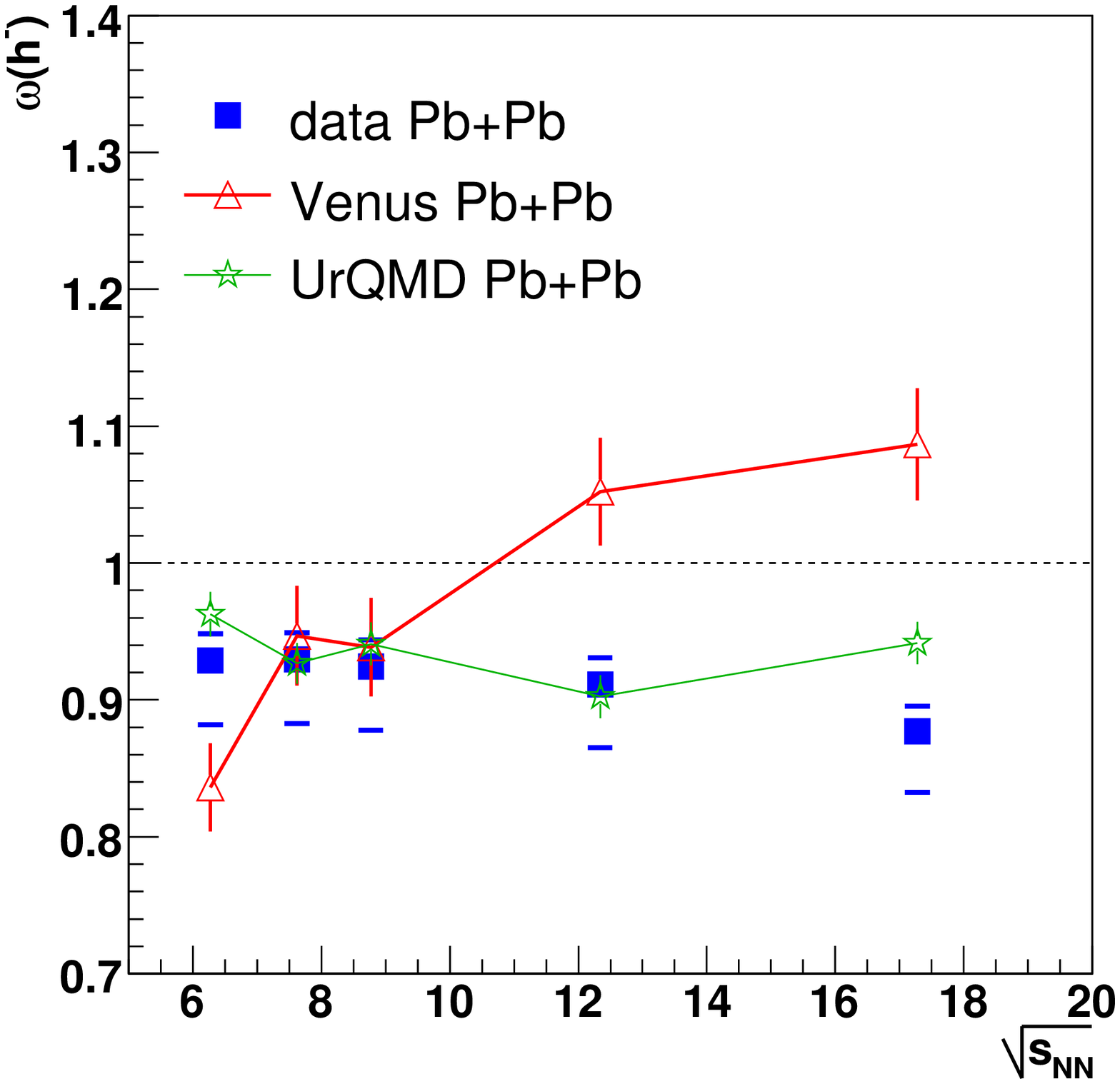}\\
\includegraphics[height=8cm]{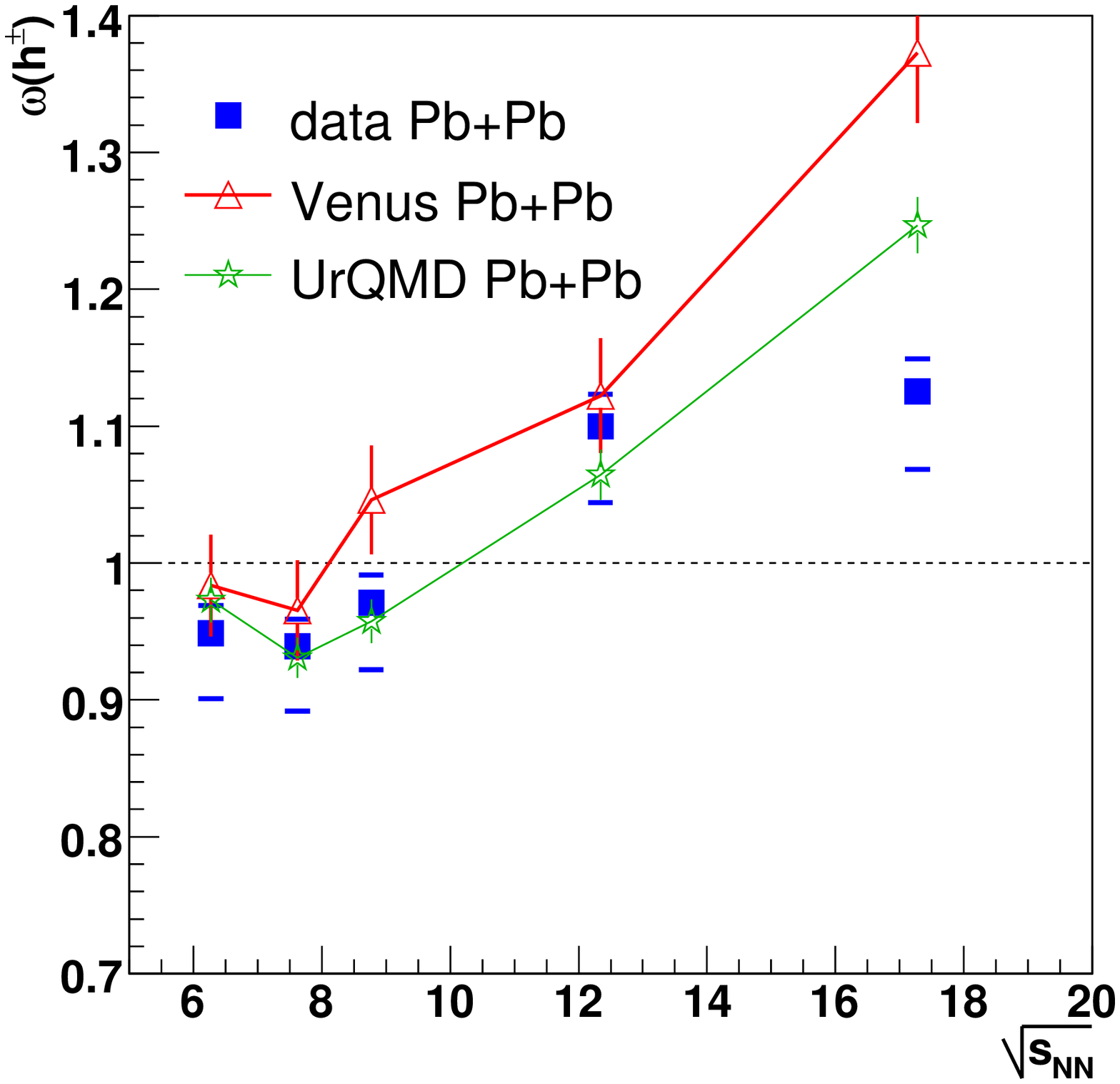}
\caption{\label{ed_strhadmod}Energy dependence of multiplicity fluctuations in the most central 
($C<1\%$, Venus calculations: $C<2\%$)
$Pb+Pb$ collisions in comparison to string hadronic 
model (Venus and UrQMD) predictions for $\omega(h^+)$ (top left),
$\omega(h^-)$ (top right) and $\omega(h^\pm)$ (bottom). Horizontal bars indicate systematic errors of data points.}
\end{figure}

The results on energy dependence of multiplicity fluctuations in $Pb+Pb$ collisions are shown in figure~\ref{ed_strhadmod}. 
It can be seen that  at all energies the scaled variance for positively
and negatively charged hadrons is smaller than $1$, the value for a Poissonian distribution.

A model independent comparison of results at different energies is not possible because the acceptance is different for each energy 
(figure~\ref{acceptance}).

The scaled variance for all charged hadrons is larger than the scaled variance for positively and negatively charged hadrons. This is probably due to the
effect of resonance decays which correlated final state hadrons and thus leads to an increase of fluctuations. This effect is largest for all charged hadrons
as the majority of resonances decay into one positively and one negatively charged daughter. It is smallest for negatively charged hadrons as only a very few resonances
decay into two negatively charged hadrons (e.g. $\Delta^{--} \rightarrow \pi^- + p^-$).

The multiplicity fluctuations of all charged hadrons in $Pb+Pb$ collisions at the top SPS energy have been studied previously by
WA98~\cite{Aggarwal:2001aa}.
A direct comparison of scaled variance of multiplicity distributions obtained in NA49 to the results of WA98
is not possible because of different experimental acceptance and centrality selection.
The WA98 results were obtained in a pseudo\-rapidity region around midrapidity ($2.35\leq \eta \leq 3.75$), whereas the NA49 results are obtained in
the forward hemisphere ($1.08 < y(\pi) < 2.57$).
The centrality of the collisions in WA98 was determined by fixing transverse energy in a calorimeter located in the pseudo\-rapidity interval 
$3.5 \leq \eta \leq 5.5$. In contrast in NA49 centrality is fixed using the energy in the forward Veto calorimeter (see section~\ref{centr_det}).

\subsection{String- Hadronic Models}

\begin{figure}
\begin{center}
\includegraphics[height=6cm]{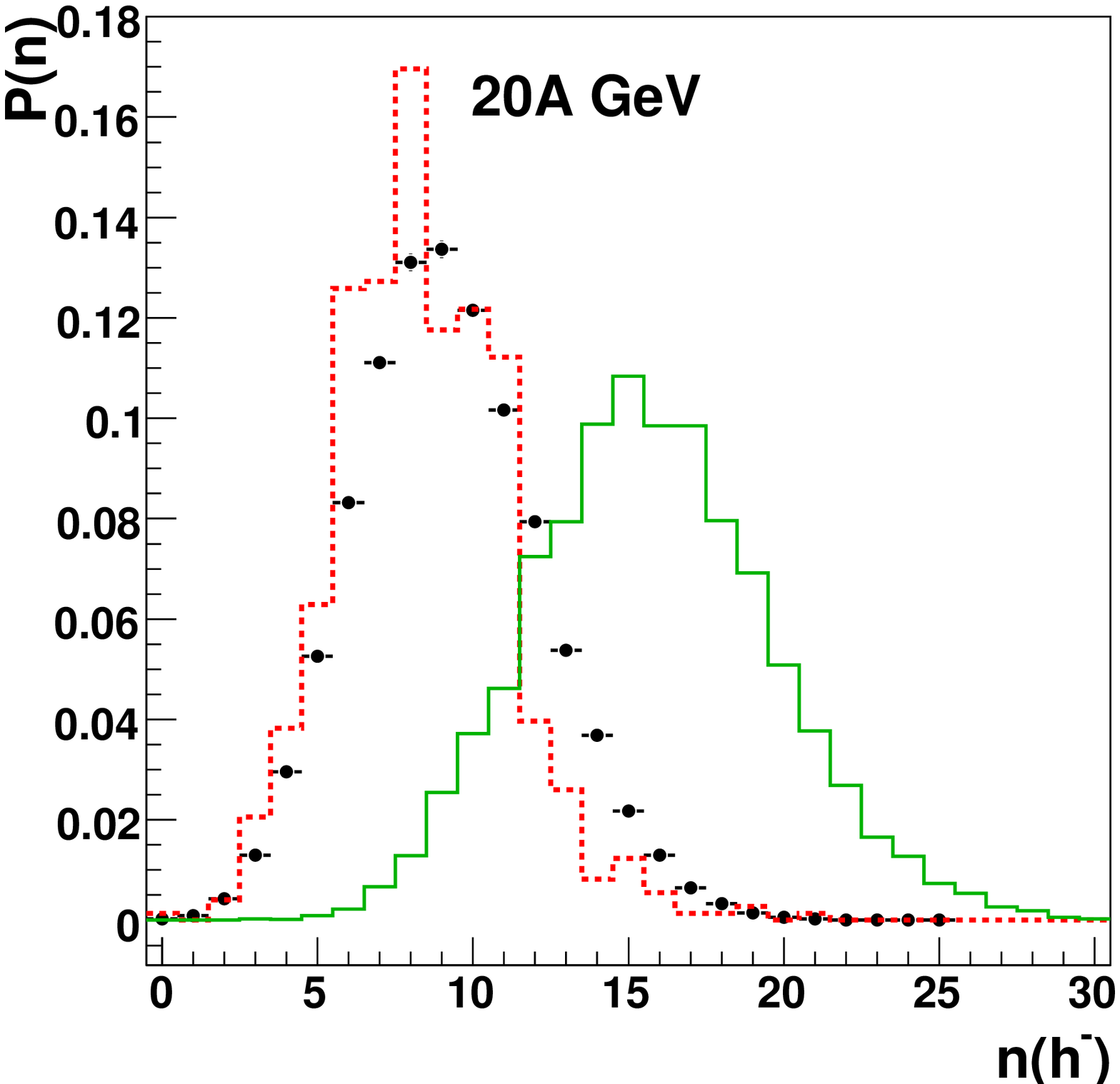}
\includegraphics[height=6cm]{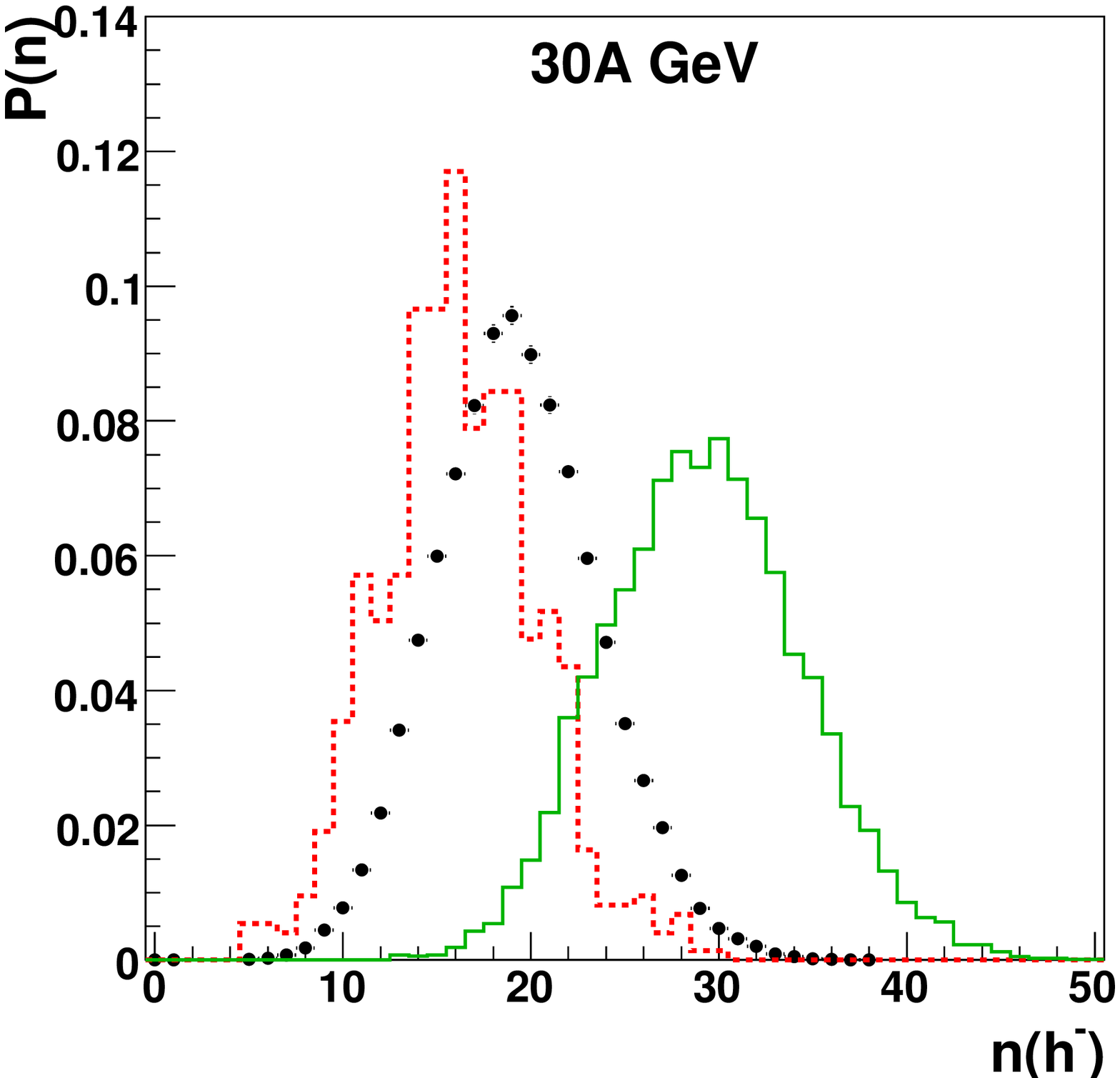}
\includegraphics[height=6cm]{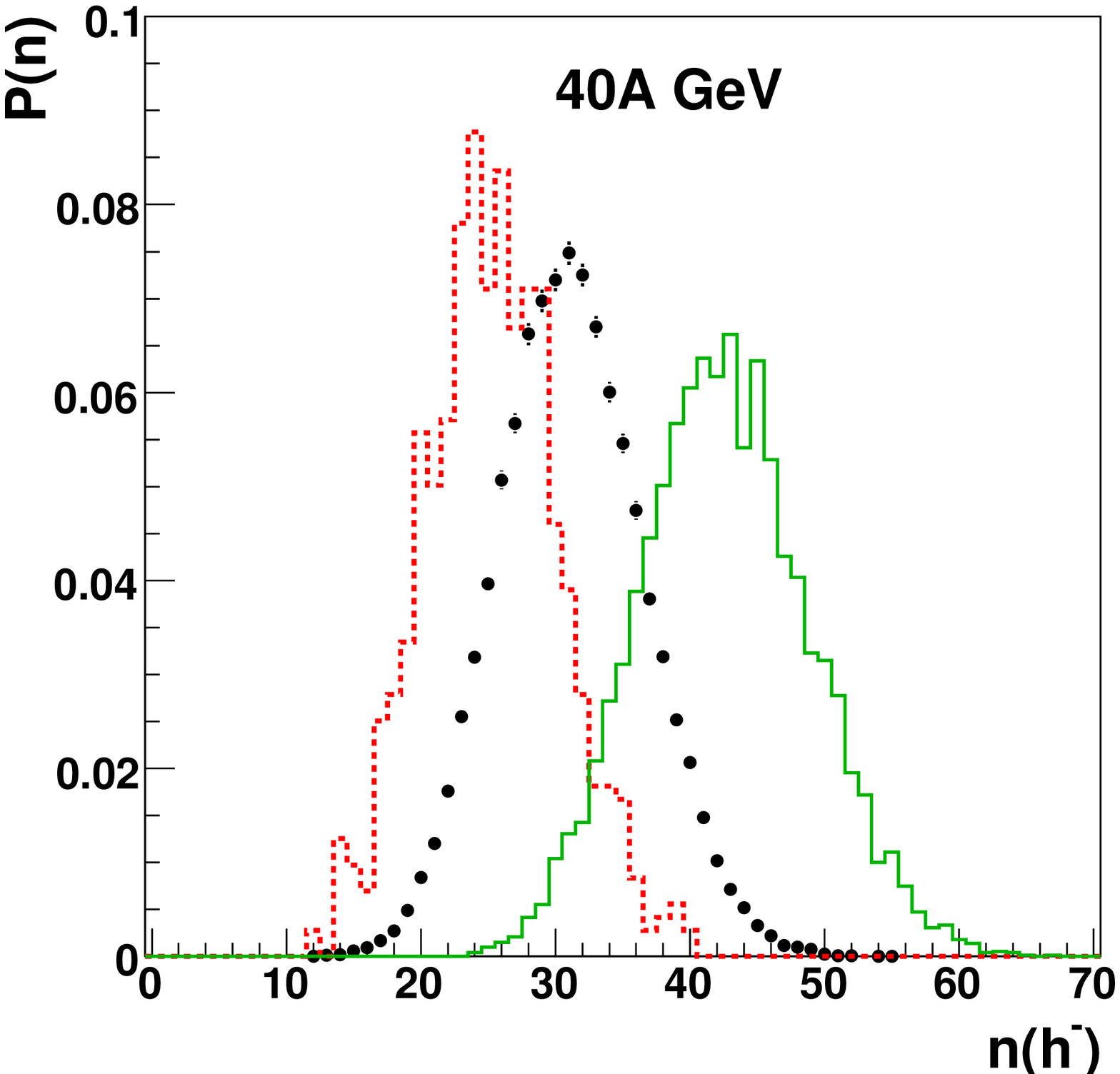}
\includegraphics[height=6cm]{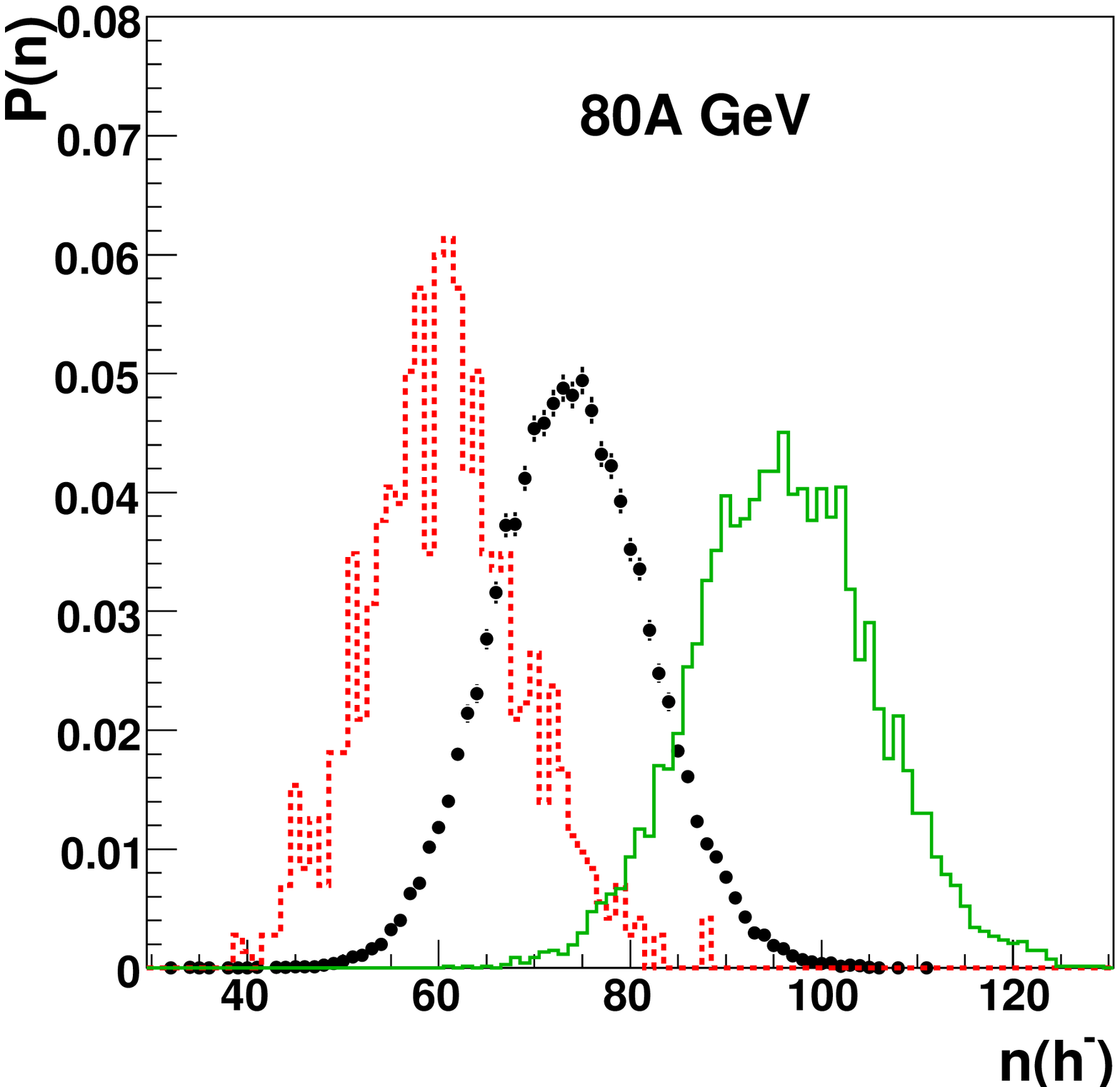}
\includegraphics[height=6cm]{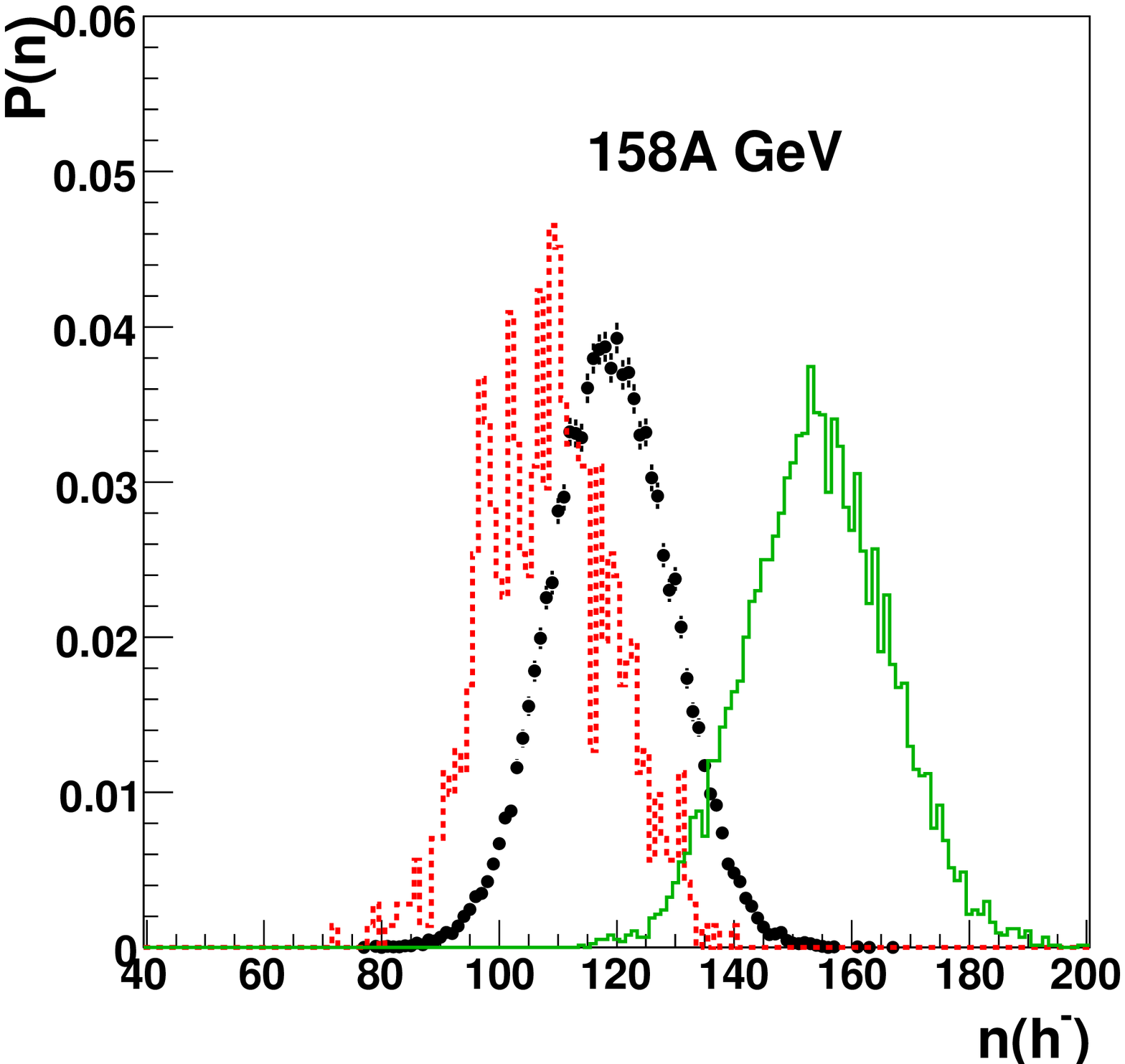}
\end{center}
\caption{\label{mult_dist_models}Multiplicity distributions of negatively charged hadrons in the most central ($C<1\%$) $Pb+Pb$ collisions from $20A$ to 
$158A$ GeV; data - black circles,
Venus model - dashed red line and UrQMD - solid green line. Note that the multiplicity fluctuations are not corrected for the finite centrality bin width.}
\end{figure}

For a comparison of data to string hadronic models (Venus $4.12$~\cite{Werner:1993uh} and UrQMD $1.3$~\cite{Bass:1998ca})
the model predictions were calculated in the data acceptance.
The centrality selection in the models is done in a similar way as in the data by a detailed simulation of 
the acceptance of the Veto Calorimeter.

The multiplicity distributions for negatively charged hadrons in data and models are compared in figure~\ref{mult_dist_models}.
In comparison to data the multiplicity distributions in Venus are shifted towards lower multiplicities, whereas in UrQMD they are
shifted towards higher multiplicities.

Figure~\ref{ed_strhadmod} shows a comparison of the scaled variance between data and the predictions of the Venus and UrQMD model. 
The Venus model 
overpredicts scaled variance for central $Pb+Pb$ collisions at most energies. In general the scaled variance obtained by the UrQMD model is in 
agreement with data. Only for 
$\omega(h^+)$ and $\omega(h^-)$ at $158A$ GeV as well as for $\omega(h^\pm)$ at $80A$ and $158A$ GeV the differences are significant.

\subsection{Acceptance Dependence}

Only a fraction of all produced particles is used for scaled variance determination in the experiment.
A simple parametrisation of the acceptance dependence of $\omega$ would be useful for comparison of different energies
and a comparison to models. In Appendix~\ref{calc_accdep} a derivation of a simple formula describing the dependence of the scaled variance on the acceptance 
is given. It is obtained under the assumption that all particles are produced
independently in momentum space.
If a fraction $p$ of all particles $N_{tot}$ is accepted, the scaled variance $\omega$ is related to the scaled variance $\omega_{tot}$ in $4\pi$ as:
$\omega=\frac{Var(N)}{<N>}=1+p \left( \omega_{tot} -1 \right)$ (see equation~\ref{accdep_svar}).
\begin{figure}
\centerline{\includegraphics[height=8cm]{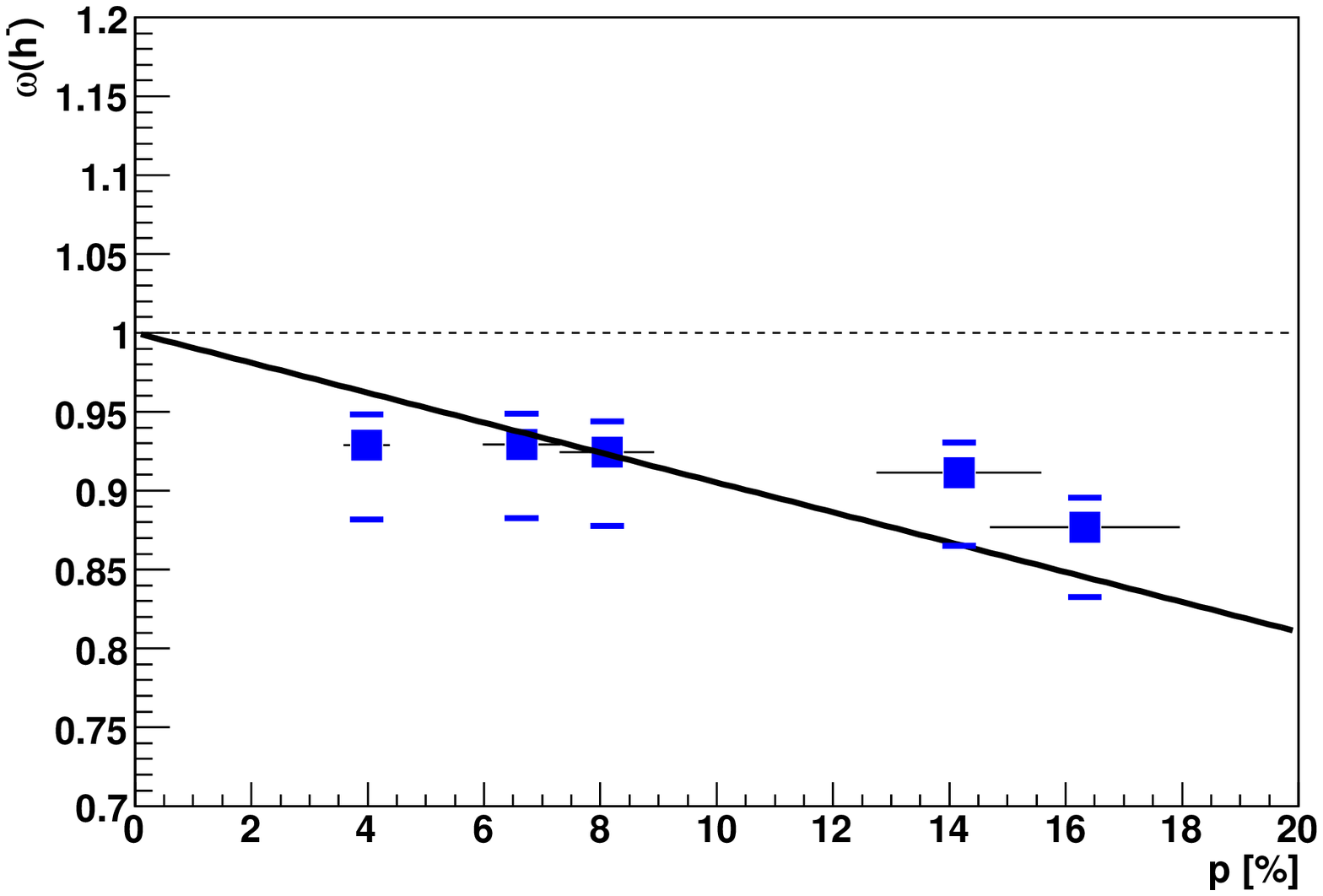}}
\caption{\label{accdep_data}Acceptance dependence of scaled variance for negatively charged hadrons. The data points corres\-pond from left to right to the
energies of $20A$ to $158A$ GeV. Horizontal bars indicate systematic errors of data points.}
\end{figure}

In figure~\ref{accdep_data} the fluctuations of negative hadrons for the five analysed energies are plotted as a function of the fraction of accepted tracks.
If the simple acceptance dependence~\ref{accdep_svar} works and if the scaled variance in $4\pi$ is similar for all energies, the data points
for all five energies should lie on a common straight line which starts at $\omega=1$ for $p=0$.
It can be seen that the data seem to follow approximately the predicted scaling. A more detailed study of rapidity and transverse momentum dependence of multiplicity
fluctuations is in progress, but first results indicate that the acceptance scaling seen in figure~\ref{accdep_data} 
should be considered only as a rough approximation.\\
For $\omega(h^+)$ this scaling does not work so well probably because a larger fraction of resonance decays (e.g. $\Delta^{++}$) into two positively charged daughters,
introducing more correlation in momentum space.
The situation is even worse for $\omega(h^\pm)$ because both daughters of a decay into two unlike sign charged hadrons (e.g. $\rho -> \pi^+ + \pi^-$)
are used for the fluctuation analysis.
Therefore acceptance scaling plots are shown only for $\omega(h^-)$.

\subsection{Statistical Models}

\begin{figure}
\centerline{\includegraphics[height=8cm]{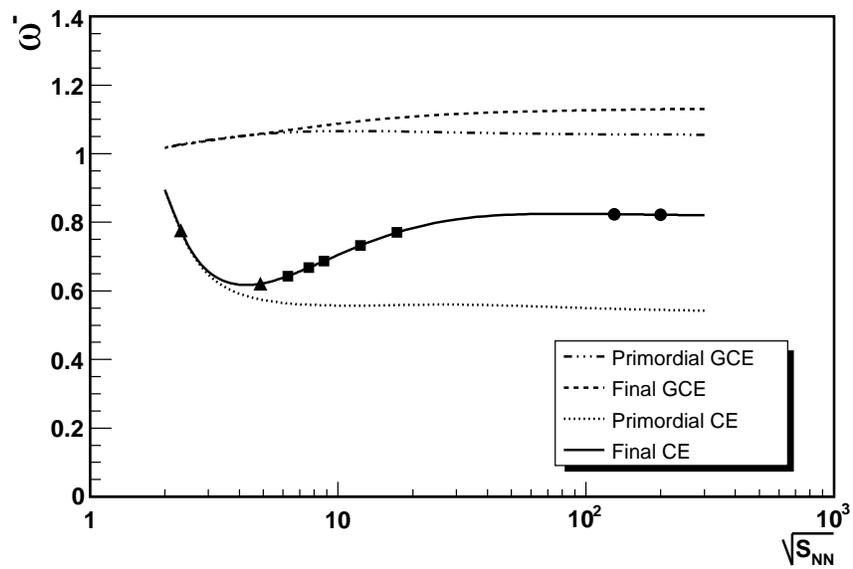}}
\caption{\label{statmod_hm}Energy dependence of the scaled variance of negatively charged hadrons in full acceptance in statistical hadron-resonance gas 
model~\cite{Begun:2006jf}
for grand-canonical (GCE) and canonical (CE) ensemble. The ``Final'' lines show results after resonance decays.}
\end{figure}

Predictions of the statistical hadron-resonance gas model~\cite{Begun:2006jf} on energy dependence of the scaled variance of negatively charged hadrons are shown
in figure~\ref{statmod_hm}. 
In this model multiplicity fluctuations in $4\pi$ were calculated for the limit of $V \rightarrow \infty$ using both canonical and grand-canonical ensembles. 
The effects of resonance decays and quantum statistics are taken into account.
The calculated value $\omega$ for the canonical ensemble is much lower than for the grand-canonical one in the SPS energy regime.
In the canonical ensemble fluctuations are
suppressed due to conservation laws of electric charge, baryon number and strangeness by about a factor of two. Preliminary calculations show a further
suppression of fluctuations in the micro-canonical ensemble due to additional energy conservation.

\begin{figure}
\centerline{\includegraphics[height=8cm]{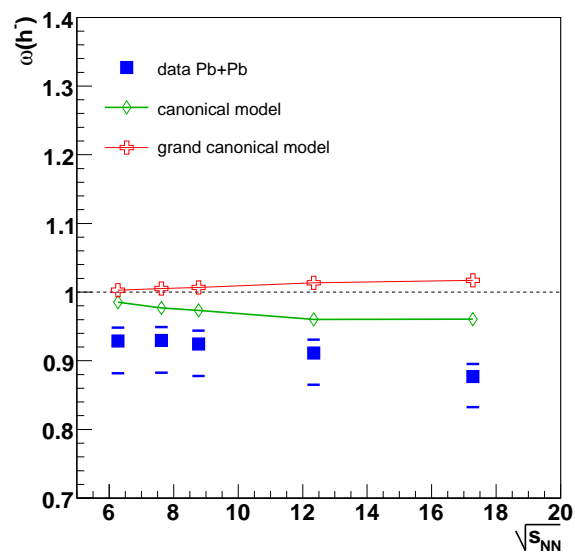}}
\caption{\label{ed_statmod}Energy dependence of multiplicity fluctuations of negatively charged hadrons in $Pb+Pb$ collisions in comparison to 
canonical and grand-canonical statistical hadron-resonance gas model~\cite{Begun:2006jf}. Horizontal bars indicate systematic errors of data points.}
\end{figure}

Figure~\ref{ed_statmod} shows data on multiplicity fluctuations in comparison to the model calculations.

The scaled variance in the limited acceptance was extrapolated using equation~\ref{accdep_svar} from the model results in full acceptance.
As discussed above equation~\ref{accdep_svar} may be considered as a reasonable approximation only for $\omega(h^-)$.
Use of equation~\ref{accdep_svar} introduces an uncertainty in the comparison of data and model. Note also that in the model calculations the volume of the fireball 
is fixed, but in data the energy in the Veto calorimeter. 

The comparison in figure~\ref{ed_statmod} shows that multiplicity fluctuations in data are much lower than predicted by the grand-canonical model.
This is an indication of suppression of fluctuations in relativistic gases due to conservation laws.
The predictions of the canonical ensemble are also higher than the data.
Energy- momentum conservation and the finite volume of hadrons are expected to cause an additional suppression of fluctuations.

\section{Summary}

Multiplicity fluctuations in the most central $Pb+Pb$ collisions at $20A$, $30A$, $40A$, $80A$ and $158A$ GeV 
were studied by NA49 at CERN SPS.
At all energies multiplicity distributions for positively and negatively charged hadrons are significantly narrower than a corresponding Poissonian distribution. 

In comparison to data the multiplicity distributions in Venus are shifted towards lower multiplicities, whereas in UrQMD they are
shifted towards higher multiplicities.
The Venus model overpredicts scaled variance for central $Pb+Pb$ collisions at most energies, whereas the scaled variance obtained by the
UrQMD model is in approximate agreement with data.

Non-monotonic 
energy dependence of fluctuations, which might be related to the onset of deconfinement or a critical point, is not observed in data.

The hadron-resonance gas model~\cite{Begun:2006jf} in grand-canonical formulation overpredicts the measured scaled variance. Introduction of the material conservation
laws (canonical model) improve agreement with the data.

The presented results may be the first experimental observation of the effect of conservation laws in relativistic gases.

\vspace{0.5cm}

\textbf{Acknowledgements:} Acknowledgements: This work was supported by the US Department of Energy
Grant DE-FG03-97ER41020/A000,
the Bundesministerium fur Bildung und Forschung, Germany, 
the Virtual Institute VI-146 of Helmholtz Gemeinschaft, Germany,
the Polish State Committee for Scientific Research (1 P03B 006 30, 1 P03B 097 29, 1 PO3B 121 29, 1 P03B 127 30),
the Hungarian Scientific Research Foundation (T032648, T032293, T043514),
the Hungarian National Science Foundation, OTKA, (F034707),
the Polish-German Foundation, the Korea Science \& Engineering Foundation (R01-2005-000-10334-0) and the Bulgarian National Science Fund (Ph-09/05).

\begin{appendix}
\section{Acceptance Dependence of Scaled Variance}\label{calc_accdep}

Let us assume that particles are emitted independently in momentum space.\\
For a given total number of produced particles $N_{tot}$, the probability to measure a number of particles $N$ in a 
fixed acceptance $p$ is a binomial distribution.
\begin{equation}
P(N|N_{tot})=P_{N_{tot}}(N)=\frac{N_{tot}!}{N! (N_{tot}-N)!} p^N (1-p)^{N_{tot}-N}
\end{equation}
Consequently the mean number of accepted particles is
\begin{equation}
<N>=N_{tot} p
\end{equation}
and the variance of the binomial distributions is
\begin{equation}
Var(N)=N_{tot} p(1-p)
\end{equation}
If one allows the total number of produced particles to vary according to distribution $P(N_{tot})$, the probability to measure a number of particles $N$ is
\begin{equation}
P(N)=\sum_{N_{tot}} P(N|N_{tot}) P(N_{tot})
\end{equation}
The mean number of particles the in acceptance is
\begin{equation}
<N>=p <N_{tot}>
\end{equation}
The variance of $N$ is given by:
\begin{equation}
\begin{split}
Var(N)&=<Var(N|N_{tot})>+Var(<N|N_{tot}>)\\
&=<Var(N|N_{tot})>+Var(p N_{tot})\\
&=<N_{tot}> p(1-p)+p^2 Var(N_{tot})
\end{split}
\end{equation}
Finally the scaled variance is
\begin{equation}\label{accdep_svar}
\omega=\frac{Var(N)}{<N>}=1+p \left( \omega_{tot} -1 \right)
\end{equation}
or 
\begin{equation}\label{accdep_svar2}
\omega_{tot}=\frac{Var(N_{tot})}{<N_{tot}>}=1+\left. \left( \omega -1 \right) \right/ p
\end{equation}
where $\omega_{tot}$ is the scaled variance of $N_{tot}$.

\end{appendix}

\end{document}